\documentclass[12pt]{article}

\usepackage{amsmath}
\usepackage{graphicx}
\usepackage{hyperref}
\usepackage[FIGTOPCAP]{subfigure}

\begin{document}

\title{Motion of the heavy symmetric top \\ when magnitudes of conserved angular momenta are different}

\author{V. Tanr{\i}verdi \\
E-mail: tanriverdivedat@googlemail.com \\
Address: Bahad{\i}n Kasabas{\i} Ceylanlar Sok. No:10 \\
66710 Sorgun / Yozgat TURKEY  
}
\date{}

\maketitle

\begin{abstract}

A symmetric top or gyroscope can start its motion with different initial values. 
One can not decide the motion type only by looking at these initial values.
For example, in a motion, precession angular velocity can be negative at the beginning, but the top's overall precession can be positive; or initially, the gyroscope spins in one direction, but in a later stage one can find it while spinning in the other direction.
On the other hand, if one knows different properties and types of motion, one can know what will happen.
In this work, we have studied the classification of motion type by using constants of motion.
We have also studied changes in dynamic variables.
We have given examples of different types of motion and solved them numerically.

\end{abstract}

\newpage

\section{Introduction}
\label{intro}

The symmetric top problem is one of the long-studied topics of physics. 
The scientific study on this topic starts with Euler and continues through Lagrange, Poinsot, Kovalevskaya and so on. 
Euler has obtained equations describing motions of rotating rigid bodies under the influence of torque in the 1750's \cite{Euler1752, Euler1765, Marquina2017}, and these equations are known as Euler equations.
These equations can result in complicated coupled equations.
Most of the time, obtaining their analytical solution is impossible, and there are various approaches to the problem.
One can find a summary of studies till the 1960s in Leimanis's work \cite{Leimanis}.

This problem is studied with geometric techniques at the beginning. 
Constants, mostly obtained using conserved quantities, are used to study the motion in these geometric approaches.
One of the important works with these geometric techniques on this problem is given by Routh \cite{Routh}, and it is one of the core works on this subject.
There is another important work on the motion of the symmetric top involving geometric techniques written by Klein and Sommerfeld \cite{KleinSommerfeld}.
Gray's work also uses geometric techniques and utilizes these two works \cite{Gray}.

Works of Lamb and Crabtree are important studies on the motion of the symmetric top \cite{Crabtree, Lamb},
and these works, together with Whittaker\& McCrae's and Synge \& Griffith's ones, provide an important step in the passage from geometric parameters to physical parameters \cite{WhittakerMcCrae, SyngeGriffith}.
In some of the later works, the constants of geometric techniques start to leave their place to physical quantities.

A cubic function, which is already used in geometric studies, can be used to determine turning angles for nutation and motion type in some cases \cite{Scarborough, MacMillan, ArnoldMaunder, Groesberg, JoseSaletan}.
The cubic function derived by using constants of motion can be used to reduce the degrees of freedom in the dissipation free case.
This kind of procedure is known as Routh reduction since Routh has given the general procedure for this type of reduction \cite{Routh2, MarsdenRatiu, LangerockAA, Capriotti}.

In some of the relatively recent works, the usage of the cubic function is modified, and effective potential together with the constants of motion is started to be used to study the motion \cite{Hauser, McCauley, MarionThornton, Symon, LandauLifshitz, Taylor}.
Both methods are equivalent, but considering the motion by using effective potential can be found more convenient since imagining inclination angle $\theta$ is easier, and the minimum of effective potential corresponds to the maximum value of nutation angular velocity \cite{Tanriverdi_ueff}.
In some works, both effective potential and the cubic function are used \cite{Arnold, Corinaldesi, MatznerShepley, Arya, Goldstein, Greiner, FowlesCassiday}.
By considering the mentioned advantages, we will use effective potential and constants of motion to study the motion.

There are also other approaches to the symmetric top problem.
One of these studies is Audin's work, and his approach can be considered as a modern version of geometric study or mathematical physics \cite{Audin}.
Leimanis, Chernousko \& Akulenko \& Leshchenko and some other authors studied the problem by using angular velocities along the axes of the body coordinate system \cite{Leimanis, ChernouskoAkulenkoLeshchenko, ge2008, amer2019}.
Some other authors considered the problem dynamically \cite{BargerOlson} or did not study different types of motion \cite{Greenwood, Gregory}.
Sussman and Wisdom studied the problem dynamically and numerically \cite{Sussman}.

After 1960, there are various papers on the motion of a gyroscope or symmetric top, some of them can be found in references, and one can learn different aspects of the motion of a symmetric top from these works \cite{Grubin1962, Prescott1963, Larmore1964, Snider1965, Case1977, Anderson1973, Whitehead1983, Panayotounakos1990, Case1992, Lewis1992, Pina1993, Soodak1994, Schonhammer1998, Springborn2000, Borisov2001, Butikov2006, Moreno2008, Provatidis2012, Kurt2012, Cross2013, Bhattacharjee2013, Zhuravlev2014, Levyraz2015, Moralesea2016, Pina2017, Tanriverdi2019, MarsdenRatiuScheurle}.
Some of these studies employ modern geometric treatments or mathematical physics, and some of them employ different physical approaches to the problem.
However, none of these works uses conserved angular momenta and energy or any other quantity related to energy to determine the motion type when the magnitudes of conserved angular momenta are different.

Even if some authors did not consider different types of motion, it is one of the essential topics in studies related to the symmetric top or gyroscope.
Different types of motion are classified according to the motion of the symmetry axis which is determined by nutation and precession angular velocities.
In Routh's work, one can find a classification which is done by using geometric parameters \cite{Routh}.
In Diemel's work, different types of motion are considered, and they are classified by using conserved angular momenta without considering possible effects of energy \cite{Diemel}.
In most of the classical mechanics' books, three different motion types together with the sleeping top and regular precession are given as examples though there are more.

Regular or steady precession is one of the important kinds of motion to understand the motion of the symmetric top.
In Routh's work, the regular precession is explained by using the cubic function.
Klein and Sommerfeld discussed the regular precession from different perspectives \cite{KleinSommerfeld, Goldstein}.
Regular precession can be seen in all different possible relations between conserved angular momenta.
The possibility of the regular precession is studied previously when conserved angular momenta are equal to each other \cite{Tanriverdi2020}.
In this work, we will study cases when magnitudes of conserved angular momenta are different.
To study this motion, we will use the derivative of effective potential like some other works \cite{MarionThornton, Symon, Arya}.
Conditions for the regular precession can also be obtained from cubic function \cite{Goldstein, FowlesCassiday} or from equations of motion \cite{Routh, ArnoldMaunder, Taylor, Corinaldesi}.
These are equivalent approaches.

In this work, we will use conserved angular momenta, a constant derived from energy, the minimum of effective potential and another constant derived from parameters of a gyroscope or symmetric top to determine the motion type.
This method is equivalent to Routh's classification in some cases.
We will study the relation between Routh's classification and classification of this work in appendix 3,
and we will show consistent and inconsistent parts of these two classifications.

There are interesting properties in the motion of the symmetric top other than the change of the symmetry axis.
Spin reversal is one of these.
In previous works, it is experimentally observed \cite{Cross2013} and studied \cite{Tanriverdi2019, Tanriverdi2020b}.
In this work, we will include a spin reversing motion while studying other possible motions, and we will study the required conditions for the spin reversal in appendix 2 which is not studied previously.

Another interesting change in precession angular velocity is also obtained, in which precession angular velocity becomes equal at both turning angles.
This is not noticed previously.

In section \ref{eom}, a review of governing equations for the motion of a spinning heavy symmetric top will be given, which is based on some well-known works \cite{Symon, LandauLifshitz, Taylor, Goldstein, Greiner, FowlesCassiday}.
In section \ref{dtm}, different types of motions will be studied by considering constants of motion, and we will give examples for each case and solve them numerically.
In section \ref{rd}, there will be a conclusion.
We will study the change in precession and spin angular velocity in appendix 1 and 2, respectively. 
As already mentioned, we will study and compare Routh's classification and the classification of this work in appendix 3.
In appendix 4, we will give the results of numerical solutions.

\section{Equations of motion for a spinning symmetric top and its general motion}
\label{eom}

A symmetric top or gyroscope can be defined with its moments of inertia $I_x=I_y$ and $I_z$, mass $M$, and the distance from the fixed point to the center of mass $l$.
These can affect the motion of the symmetric top or gyroscope, and we will mention their effects at some points.

\begin{figure}[!h]
	\begin{center}
		\includegraphics[width=7cm]{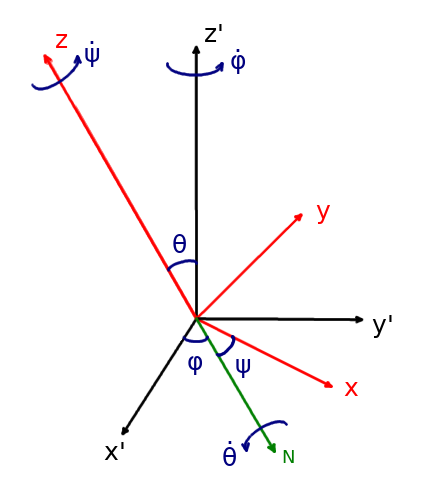}
		\caption{Stationary reference frame ($x', y', z'$), body reference frame ($x, y, z$), line of nodes $N$, Euler angles ($\theta, \phi, \psi$) and angular velocities ($\dot \theta, \dot \phi, \dot \psi$).}
		\label{fig:reffrs}
	\end{center}
\end{figure}

Euler angles ($\theta(t)$, $\phi(t)$ and $\psi(t)$), corresponding angular velocities ($\dot \theta(t)$, $\dot \phi(t)$ and $\dot \psi(t)$) and time $t$ are 7 variables that can be used to define the motion of a spinning symmetric top.
One can see angles and angular velocities in figure \ref{fig:reffrs}.
For a spinning symmetric top under the gravitational force, in the dissipation free case, Lagrangian is \cite{Goldstein}
\begin{equation}
	L=\frac{I_x}{2}(\dot \theta ^2 + \dot \phi ^2 \sin^2 \theta)+\frac{I_z}{2}(\dot \psi+\dot \phi \cos \theta)^2-M g l \cos \theta, 
	\label{lagrngn}
\end{equation}
where $\theta$ is the angle between body $z$-axis and stationary $z'$-axis, $\dot \phi$ defines rotation around the stationary $z'$-axis, $\dot \psi$ defines rotation around the body $z$-axis and $\dot \theta$ defines rotation around the line of nodes, which is the intersection of stationary $x'y'$-plane and body $xy$-plane.
$\dot \phi$, $\dot \psi$ and $\dot \theta$ are known as precession, spin and nutation angular velocities, respectively.
We should note that $g$ is the gravitational acceleration.

$\theta$ can have values between $0$ and $\pi$.
When $\theta<\pi/2$, one can imagine a symmetric top spinning on the ground, 
and when $\pi/2<\theta$, it can be considered as suspended from its fixed point while spinning.

By using Euler-Lagrange equation related to $\theta$, the angular acceleration $\ddot \theta$ can be obtained as
\begin{equation}
	\ddot \theta=\frac{\sin \theta}{I_x} \left[ Mgl+I_x \dot \phi^2 \cos \theta-I_z\dot \phi^2 \cos \theta-I_z \dot \phi \dot \psi \right]. \label{ddottheta}
\end{equation}
Form other two Euler-Lagrange equations, angular accelerations $\ddot \phi$ and $\ddot \psi$ can be obtained as
\begin{eqnarray}
	\ddot \phi&=& \frac{\dot \theta}{I_x \sin \theta}\left[ I_z \dot \psi +I_z \dot \phi \cos \theta -2 I_x \dot \phi \cos \theta \right],  \label{ddotphi} \\
	\ddot \psi&=& -\frac{\cot \theta}{I_x} \left[ I_z \dot \theta \dot \psi +I_z \dot \theta \dot \phi \cos \theta -2 I_x \dot \theta \dot \phi \cos \theta \right]+\dot \theta \dot \phi \sin \theta.  \label{ddotpsi}
\end{eqnarray}
These three equations are coupled and can be solved numerically. 
However, these equations do not give much insight into the motion.
To get insight, we should consider the constants of motion and effective potential.

From equation \eqref{lagrngn}, it can be seen that $t$, $\psi$ and $\phi$ are not present in Lagrangian, and according to Noether's theorem, there are three constants of motion: energy $E$, angular momentum along body $z$-axis $L_{z}$, and angular momentum along stationary $z'$-axis $L_{z'}$, respectively.
Two conserved angular momenta can be obtained by using two of Euler-Lagrange equations as \cite{Greiner}
\begin{eqnarray}
	L_{z}&=&I_z(\dot \psi+\dot \phi \cos \theta), \nonumber \\
	L_{z'}&=&I_x \dot\phi \sin^2 \theta +I_z (\dot \psi+\dot \phi \cos \theta)\cos \theta .
	\label{angmom}
\end{eqnarray}
In some works, these conserved angular momenta are shown by $p_\psi$ and $p_\phi$ since they are canonical momenta conjugate to $\psi$ and $\phi$, respectively \cite{Symon, Goldstein}.
These equations show that $L_z$ and $L_{z'}$ are conserved separately.
Both $\dot \phi$ and $\dot \psi$ are present in both equations, and 
this shows that any change in one of these angular velocities will result in a change in the other one.

By using the two conserved angular momenta, one can define $a$ and $b$ as $a=L_z/I_x$ and $b=L_{z'}/I_x$ \cite{Goldstein}.
And, naturally, these two parameters correspond to conserved angular momenta.
Since $L_z$ and $L_{z'}$ are conserved separately so do $a$ and $b$.
We should note that $a$ and $b$ can have negative values, and their dimension is the same as angular velocity.

Energy is another constant of motion.
For the spinning heavy symmetric top, it can be written as \cite{Goldstein}
\begin{equation}
	E=\frac{I_x}{2}(\dot \theta ^2 + \dot \phi ^2 \sin^2 \theta)+\frac{I_z}{2}(\dot \psi+\dot \phi \cos \theta)^2+M g l \cos \theta.
	\label{enrgy}
\end{equation}
By using constancy of $L_z$, it is possible to define another constant corresponding to the energy in terms of $a$ and $b$ as \cite{Goldstein}
\begin{equation}
	E'=\frac{I_x}{2}\dot \theta ^2 +\frac{I_x}{2} \frac{(b-a \cos \theta)^2}{\sin^2 \theta}+M g l \cos \theta. \label{eprime}
\end{equation}
This constant can be used to determine the motion type of the symmetric top or gyroscope and helpful to understand the motion.

By using equation \eqref{eprime} and change of variable $u=\cos \theta$, one can find $\dot u^2=f(u)$, where $f(u)$ is the cubic function mentioned in the introduction and can be written as  \cite{Goldstein}
\begin{equation}
	f(u)=(\alpha-\beta u)(1-u^2)-(b-a u)^2, 
	\label{cbf}
\end{equation}
where $\alpha=2E'/I_x$ and $\beta =2 Mgl/I_x$.
For proofs, the usage of the cubic function is better.

\begin{figure}[!h]
	\begin{center}
		\includegraphics[width=7cm]{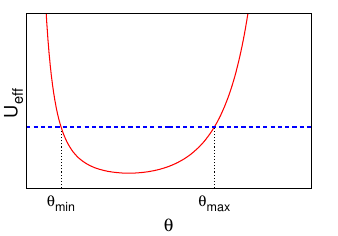}
		\caption{General structure of $U_{eff}$ (red curve) with respect to $\theta$, and $E'$ (dashed blue line).
		Intersection points of $E'$ and $U_{eff}$ give turning angles; $\theta_{min}$ and $\theta_{max}$.
		}
		\label{fig:ueffg}
	\end{center}
\end{figure}

But to utilize the advantage of imagination, we will consider the effective potential, and it can be defined as \cite{Goldstein}
\begin{equation}
	U_{eff}(\theta)= \frac{I_x}{2}\frac{(b-a \cos \theta)^2}{\sin^2 \theta}+Mgl \cos \theta.
	\label{ueff}
\end{equation}
In this work, we will consider the situation when $|a|\ne|b|$.
This effective potential will go to infinity at the domain boundaries of $\theta$, $[0,\pi]$, and have a minimum between these boundaries.
Then, the form of effective potential is like a well.
The general structure of $U_{eff}$ can be seen in figure \ref{fig:ueffg}.
There is an asymmetry in the form of the effective potential \cite{Tanriverdi_ueff}.
The minimum of the effective potential can be negative, and its position and magnitude depend on $I_x$, $Mgl$, $b$ and $a$.
In figure \ref{fig:ueffg}, $E'$ is also shown.
$E'$ can not be smaller than the minimum of the effective potential according to their definitions.

When $E'$ is equal to the minimum of $U_{eff}$, only one $\theta$ value is possible.
Since $\theta$ does not change, $\dot \theta$ is always zero, and $\dot \phi$ and $\dot \psi$ do not change throughout the motion.
This type of motion is known as the regular precession or steady precession.
One can use the precession period to understand the regular precession quantitatively and qualitatively by considering the motion in $\phi$ from $0$ to $2 \pi$.

When $E'$ is grater than the minimum of $U_{eff}$, similar to figure \ref{fig:ueffg}, the intersection points of $E'$ and $U_{eff}$ correspond to extremum values of $\theta$; $\theta_{min}$ and $\theta_{max}$.
If the initial value of $\theta$ is equal to $\theta_{min}$, as time passes, $\theta$ increases till reaching $\theta_{max}$ and then starts to decrease.
This decrease continues till reaching $\theta_{min}$ \cite{Taylor, FowlesCassiday}.
Therefore, the extremum values of $\theta$ can be considered as turning angles.

The mentioned change in $\theta$ periodically repeats itself.
The period in the motion from $\theta_{min}$ to $\theta_{min}$ or from $\theta_{max}$ to $\theta_{max}$ can be defined as nutation period, which is helpful in calculations and understanding.
For the regular precession, the nutation period is meaningless since there is not any change in $\theta$.
Nevertheless, one can use the precession and nutation periods by taking into account their assistance in understanding and analyzing.

The periodic change in $\theta$ takes place together with a periodic change in $\dot \theta$.
$\dot \theta$ has a positive sign while $\theta$ is increasing, and vice versa.
And, its magnitude depends on the difference between $E'$ and $U_{eff}(\theta)$.
The asymmetry in $U_{eff}$ results in an asymmetry in the graph of $\dot \theta$.

Both $\dot \phi$ and $\dot \psi$ are present in both conserved angular momenta, given by equations \eqref{angmom}, and
changes in $\dot \phi$ and $\dot \psi$ obey conservation of these angular momenta.
$\dot \phi$ can be obtained in terms of $a$ and $b$ as a function of $\theta$ \cite{Goldstein}
\begin{equation}
	\dot \phi(\theta)=\frac{b - a \cos \theta}{\sin^2 \theta}. \label{phidot}
\end{equation}
This equation shows that $\dot \phi$ changes as $\theta$ changes. 
Since $a$ and $b$ are specified by initial values and they do not change in the dissipation free case, $\dot \phi$ can be considered as a function of $\theta$ alone.
We have seen that $\theta$ takes values between its two turning angles periodically, then, the change in $\dot \phi$ is also periodic, and the change in $\dot \phi$ from $\theta_{min}$ to $\theta_{max}$ is reverse of its change from $\theta_{max}$ to $\theta_{min}$.

An interesting situation takes place when $U_{eff}$ becomes equal to $Mglb/a$, and in this case by using equation \eqref{ueff}, one can write
\begin{equation}
	Mgl\frac{b}{a}=\frac{I_x}{2}\frac{(b-a \cos \theta)^2}{\sin^2 \theta}+Mgl \cos \theta,
	\label{ueffmglboa}
\end{equation}
and by taking $Mgl \cos \theta$ to the left-hand side and using equation \eqref{phidot}, it can be written that
\begin{equation}
	\dot \phi^2= \frac{2Mgl}{I_x a} \dot \phi.
	\label{mglboadotphi}
\end{equation}
This shows that when $U_{eff}=Mglb/a$, $\dot \phi$ is equal to either $0$ or $2Mgl /(I_x a)$.
The reverse of this statement also holds.
If $E' > Mglb/a$, $U_{eff}$ becomes equal to $Mglb/a$ at two different $\theta$ values between $\theta_{min}$ and $\theta_{max}$ provided that $Mglb/a>U_{eff_{min}}$.
It is shown in appendix 1 that when $|a|>|b|$, $\dot \phi$ can be equal to $0$ and it is either a decreasing or an increasing function, and when $|b|>|a|$, $\dot \phi$ can not be equal to $0$ and it is a decreasing function in some interval and an increasing function in the remaining interval.
These show that when $|a|>|b|$, $\dot \phi$ is equal to $0$ at one of these two $\theta$ values and it is equal to $2Mgl /(I_x a)$ at the other one; and when $|b|>|a|$, $\dot \phi$ is equal to $2Mgl /(I_x a)$ at both of these $\theta$ values.
The relation between $E'$ and $Mglb/a$ can be used to determine the motion type, and we will study different possibilities of this relation with examples.

$\dot \psi$ can be obtained as \cite{Goldstein}
\begin{equation}
	\dot \psi(\theta)=\frac{I_x}{I_z}a-\frac{b - a \cos \theta}{\sin^2 \theta}\cos\theta. \label{psidot}
\end{equation}
The change in $\dot \psi$ is also periodic since it can also be considered as a function of $\theta$ similar to $\dot \phi$, and the change in $\dot \psi$ from $\theta_{min}$ to $\theta_{max}$ will be reverse of the change in it from $\theta_{max}$ to $\theta_{min}$.
The sign change of $\dot \psi$ is also possible \cite{Cross2013, Tanriverdi2019, Tanriverdi2020b} when $|b|>|a|$, and the angle where spin reversal occurs can be found as
\begin{equation}
	\theta_{r_{1,2}}=\arccos\left[ \frac{ \frac{b}{a}\pm\sqrt{ \frac{b^2}{a^2}-4 \frac{I_x}{I_z}\left(1-\frac{I_x}{I_z}\right)} }{2\left(1-\frac{I_x}{I_z} \right)}\right].
	\label{rootdotpsi}
\end{equation}
One can find details of the change in $\dot \psi$ in appendix 2.

One can draw three-dimensional figures representing the motion of the symmetry axis of the top.
These figures are known as shapes for the locus of the figure axis or symmetry axis \cite{Goldstein}.
These figures are obtained by using two Euler angles $\theta$ and $\phi$, and their change takes place according to $\dot \theta$ and $\dot \phi$.
In general, noticeable changes in these figures occur during the sign change of these two angular velocities. 
On the other hand, the sign change in $\dot \psi$ is also important, but it does not affect shapes for the locus.

\section{Different types of motion and their solution}
\label{dtm}

In this section, we will study different types of motion by considering relations between $a$ \& $b$ and $E'$ \& $U_{eff_{min}}$ or $E'$ \& $Mglb/a$ and solve them numerically with two methods.
The first numerical solution is obtained by integrating angular accelerations $\ddot \theta$, $\ddot \phi$ and $\ddot \psi$ which are given by equations \eqref{ddottheta}, \eqref{ddotphi} and \eqref{ddotpsi}.
One can get $\dot \theta(t)$, $\dot \phi(t)$ and $\dot \psi(t)$ from these integrations, and with another integration one can get also $\theta(t)$, $\phi(t)$ and $\psi(t)$.
In the second numerical solution, constants of motion are used.
The constants of motion can be obtained using initial values $\theta_0$, $\dot \theta_0$, $\dot \phi_0$ and $\dot \psi_0$.
From equation \eqref{eprime} or \eqref{cbf}, it can be seen that $\dot \theta$ can be obtained in terms of constants of motion and $\theta$, 
then one can numerically obtain $\theta(t)$ and $\dot \theta(t)$ by using one of these equations.
After obtaining $\theta(t)$, it is possible to obtain $\dot \phi(t)$ and $\dot \psi(t)$ by using equations \eqref{phidot} and \eqref{psidot}. 
The details of this numerical solution can be found in various works \cite{McCauley, MarionThornton, Symon, Arya, Tanriverdi2019}.

We should note that the integration of angular accelerations, the first method, has some advantages over the usage of constants and reduction, the second method.
In the first method, one can get results for evenly spaced time intervals.
The second method requires either specifying turning angles or using a special function in numerical code to detect turning angles.
While using equation \eqref{eprime} or \eqref{cbf} for the second method, one should choose a sign for $\dot \theta$ at turning angles.
On the other hand, numerically integrating angular accelerations does not need such specifications.
Nevertheless, we will use both methods which can be considered as a checkpoint for numerical solutions.

In the following part, $Mgl$ will be considered as positive unless otherwise is mentioned.

In this work, a symmetric top or gyroscope with parameters $I_x=I_y=22.8 \times 10^{-5} \,kg \,m^2$, $I_z=5.72 \times 10^{-5} \,kg \,m^2 $ and $Mgl=0.068 \,J$ will be considered for numerical solutions of examples similar to previous works \cite{Tanriverdi2020, Tanriverdi2019}.

\subsection{Different types of motion when $|a|>|b|$}

When $|a|>|b|$, there are four possible motions corresponding to four different values of $E'$: $E'=U_{eff_{min}}$, $E'=Mglb/a$, $E'<Mglb/a$ and $E'>Mglb/a$.
We should note that $Mglb/a$ is always greater than the minimum of $U_{eff}$ when $|a|>|b|$ \cite{Tanriverdi_ueff}.
In this section, we will study and give examples of these four possibilities.
We should also note that spin reversal is not possible when $|a|>|b|$, and details can be found in appendix 2.
When $|a|>|b|$, the overall precession direction has the same sign as $a$ \cite{Leimanis} or $\dot \psi$ since it has the same sign as $a$ in this condition (see appendix 2).

\subsubsection{Regular precession}
\label{regpr1}

We already mentioned that effective potential has a minimum.
In the regular precession, $E'$ is equal to that minimum value.

In this type of motion, the heavy symmetric top spinning with $\dot \psi_0$ should start its motion with an initial inclination angle $\theta_0$ and precession angular velocity $\dot \phi_0$ and continue its precession without nutation, $\dot \theta=0$ \cite{Goldstein}.

One can take the derivative of equation \eqref{ueff} with respect to $\theta$ and equate it to $0$ to find the minimum of $U_{eff}$.
After writing $a$ and $b$ in terms of $\dot \phi$ and $\dot \psi$ at the derivative of $U_{eff}$, one can get \cite{ArnoldMaunder}
\begin{equation}
	[\dot \phi^2 \cos \theta (I_z-I_x)+\dot \phi \dot \psi I_z-Mgl]\sin \theta =0.
	\label{dueff1}
\end{equation}
$\sin \theta$ is equal to $0$ at $\theta=0$ and $\theta=\pi$, which give infinities of effective potential.
If $\theta=\pi/2$, then this equation reduces to first order and it has only one solution $\dot \phi=Mgl/(I_z \dot \psi)$ \cite{ArnoldMaunder}.
If $I_z=I_x$, the same solution is obtained without this angle restriction, but it is not a symmetric top anymore but a spherical top. 

For a symmetric top ($\theta \ne \pi/2$), after simplifying $\sin \theta$, one can get 
\begin{equation}
	\dot \phi^2 \cos \theta (I_z-I_x)+\dot \phi \dot \psi I_z-Mgl =0.
	\label{dueff2}
\end{equation}
This equation is quadratic in $\dot \phi$, and its discriminant is $D=(I_z \dot \psi)^2+4(I_z-I_x) Mgl \cos \theta$.
If $D<0$, then the regular precession is impossible.
This means that to get a regular precession for a specified angle $\theta$, there can be a lower bound for $|\dot \psi|$; and to get a regular precession for a specified spin angular velocity $\dot \psi$, there can be a lower bound for $\theta$ when $I_x>I_z$, and an upper bound for $\theta$ when $I_z>I_x$ depending on the magnitude of $\dot \psi$.
If these are not specified, then one can always find a configuration for the regular precession:
$\dot \psi^2$ is a positive quantity and $4(I_z-I_x) Mgl \cos \theta/I_z^2$ can be positive when either $\theta \in (0,\pi/2)$ or $\theta \in (\pi/2,\pi)$,
and this shows that the regular precession is always possible as long as $\dot \psi^2>0$.
If $\dot \psi$ is equal to $0$, then this case is not the topic of motion of the spinning heavy symmetric top but the topic of conical pendula \cite{ArnoldMaunder}.

If $D=0$, there is only one $\dot \phi$ value giving the regular precession,
which is given by $\dot \phi=-I_z \dot \psi/[2 (I_z-I_x)\cos \theta]$ or $\dot \phi= 2 Mgl/(I_z \dot \psi)$ provided that $\cos \theta=-(I_z \dot \psi)^2/[4(I_z-I_x) Mgl]$ which should be satisfied for both alternatives.
If $I_z< I_x$, then $\theta$ should be between $0$ and $\pi/2$;
and if $I_z> I_x$, then $\theta$ should be between $\pi/2$ and $\pi$ to get the regular precession with single $\dot \phi$.
For the regular precession with single $\dot \phi$, magnitude of the spin angular velocity $\dot \psi$ should be smaller than $\sqrt{4 Mgl |I_z-I_x|/I_z^2}$.
Then, one can say that there are always two possible $\dot \phi$ values for regular precession when $|\dot \psi|$ is large enough.

When $D>0$, precession angular velocities for the regular precession can be found by using equation \eqref{dueff2} as
\begin{equation}
	\dot \phi_{1,2}=\frac{-I_z \dot \psi \pm \sqrt{D}}{2 (I_z -I_x) \cos \theta}.
	\label{regular12}
\end{equation}
If $\pi/2>\theta$ and $(I_z-I_x)>0$, then the roots have different signs; and if $\pi/2>\theta$ and $(I_z-I_x)<0$, then the roots have the same sign.
If $\theta>\pi/2$ and $(I_z-I_x)>0$, then the roots have the same sign; and if $\theta>\pi/2$ and $(I_z-I_x)<0$, then the roots have different signs.

There is an alternative scheme to the abovementioned procedure, and it is used in some of the works.
One can write equation \eqref{dueff2} as
\begin{equation}
	\dot \phi^2 \cos \theta-\dot \phi a+\frac{Mgl}{I_x} =0.
	\label{dueff3}
\end{equation}
One can see that the reduction of equation \eqref{dueff1} to the first order when $I_z=I_x$ is invisible in this form though the reduction when $\theta=\pi/2$ is visible.
In this case, it can be seen that the discriminant can be obtained as $\tilde D= a^2-4 Mgl \cos \theta/I_x$, and precession angular velocities giving regular precession can be obtained by using
\begin{equation}
	\dot \phi_{1,2}=\frac{a \pm \sqrt{ \tilde D}}{2 \cos \theta}.
	\label{regular12s}
\end{equation}
One can define a constant as $\tilde a=\sqrt{4 Mgl/I_x}$.
If $|a|>\tilde a$ ($|a|<\tilde a$), then the symmetric top can be designated as "strong top" ("weak top") \cite{KleinSommerfeld} related to the response of the top to torque or disturbance.
This kind of separation is useful to see the change in effective potential when $|b|=|a|$ \cite{Symon, Tanriverdi2020, Case1992}, it can be helpful when $|a|>|b|$, and it is not adaquate when $|b|>|a|$ \cite{Tanriverdi_ueff}.

Equation \eqref{dueff3} is simpler with respect to equation \eqref{dueff2}.
But there is a drawback, and one should be careful while using it since $a$ is obtained by using $\dot \phi_0$.
According to discriminant of equation \eqref{dueff3}, $\theta$ should be between $0$ and $\pi/2$ to get the regular precession with single $\dot \phi$ which is possible only for weak top.
During the analyzes of equation \eqref{dueff2}, we have seen that the regular precession with single $\dot \phi$ is still possible when $\pi/2<\theta$.
The usage of the constant $a$ shadowed that possibility.

There are also other possible shadowings.
If $\tilde D$ is greater than zero, there are two possible $\dot \phi$ values giving regular precession.
For "weak top", $\theta$ should be greater than $\pi/2$ and regular precessions take place in different directions.
For "strong top", $\theta$ can have any values and regular precessions take place in the same direction when $\theta<\pi/2$, and they take place in different directions when $\theta>\pi/2$.
From equation \eqref{regular12}, one can see that regular precession is possible for "weak top" when $\theta<\pi/2$ provided that $(I_z-I_x)>0$; and regular precessions at opposite directions are possible when $\theta<\pi/2$ provided that $(I_z-I_x)>0$, and regular precessions at the same direction are possible when $\theta>\pi/2$ provided that $(I_z-I_x)>0$

In equations \eqref{regular12} and \eqref{regular12s}, specification of $\theta$ is necessary to find $\dot \phi$ giving the regular precession.
On the other hand, for specified values of $E'$, $a$ and $b$, one can also find the angle $\theta$ giving the regular precession as long as $E'=U_{eff_{min}}$.
The cubic function, equation \eqref{cbf}, has a double root for the regular precession, and this double root can be found as $u=[ 16 Mgl E' -Mgl I_x(9 b^2+a^2) +I_x a b(2 E'+I_x a^2)]/[(2E'+I_x a^2)^2+12 Mgl( Mgl- I_x a b)]$.  
Then, one can find $\theta$ to get the regular precession for specified constants by changing the variable back.

If $U_{eff_{min}}$ is not known, then by considering $\ddot \theta=0$ and writing $\dot \phi$ and $\dot \psi$ in terms of $a$ and $b$ in this equation with change of variable $u=\cos \theta$, one can obtain $u^4- A u^2 +B u -C=0$, where $A=I_x a b/Mgl+2$, $B=I_x (a^2+b^2)/Mgl$ and $C=I_x a b/Mgl-1$. By solving this quartic equation and considering $\theta=\arccos u$ for the root which is between $-1$ and $1$, one can get the angle where effective potential is minimum. 

In this work, one of the solutions given by equation \eqref{regular12} will be used as an example for the regular precession when $|a|>|b|$; and the other one will be used as an example for the regular precession when $|b|>|a|$.
We should note that different roots do not always satisfy different relations between $a$ and $b$, sometimes they satisfy the same relation.

\begin{figure}[!h]
	\begin{center}
		\subfigure[]{
			\includegraphics[width=4.2cm]{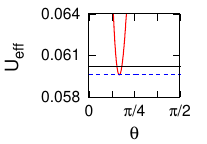}
		}
		\subfigure[]{
			\includegraphics[width=4.2cm]{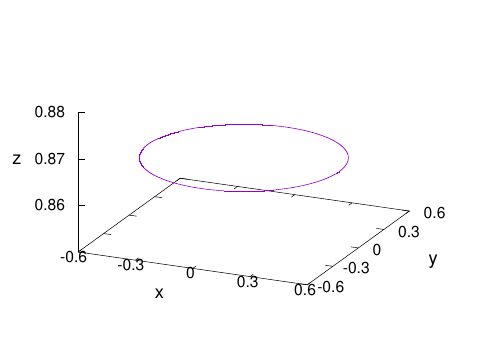}
		}
		\caption{(a) $U_{eff}$ (red curve), $E'$ (dashed blue line) and $Mgl b/a$ (black line).
		(b) Shapes for the locus on the unit sphere for the regular precession when $a>b$. 
		Initial values are $\theta_0=0.524\,rad$, $\dot \theta_0=0$, $\dot \phi_0=5.015 \,rad \,s^{-1}$ and $\dot \psi_0=250 \,rad \,s^{-1}$, and constants are $a=63.8 \,rad \,s^{-1}$, $b=56.5 \,rad \,s^{-1}$, $E'=0.0596 \,J$ and $Mgl b/a=0.0602 \,J$.
		The animated plot is available at \href{https://youtu.be/hNETGJgoG_4}{https://youtu.be/hNETGJgoG\_4}.
		}
		\label{fig:uefftt_7a1}
	\end{center}
\end{figure}

If we take $\theta_0=0.524 \,rad$ and $\dot \psi_0=250 \,rad\, s^{-1}$, one can find the root satisfying $a>b$ relation as $\dot \phi=5.015 \,rad\, s^{-1}$ for the regular precession.
In this case, $a=63.8 \,rad \,s^{-1}$, $b=56.5 \,rad \,s^{-1}$, $E'=0.0596 \,J$ and $Mgl b/a=0.0602 \,J$.
In figure \ref{fig:uefftt_7a1}(a), effective potential together with $E'$ can be seen, and one can see that $E'$ is equal to the minimum of $U_{eff}$.

In this case, one can not use the second numerical solution, since $\theta$ is constant.
However, one can integrate numerically angular accelerations, and results are available in figure \ref{fig:ttdfd_7a1} in appendix 4.
At there, it can be seen that $\theta$, $\dot \phi$ and $\dot \psi$ do not change and $\dot \theta$ is equal to $0$ throughout the motion.
In figure \ref{fig:uefftt_7a1}(b), we see shapes for the locus for the regular precession.

\subsubsection{Motion with cusps}
\label{mwc}

For motion with cusps, $E'$ is equal to $Mgl b/a$.
We have seen that when $E'=Mgl b/a$ and $|a|>|b|$, $U_{eff}$ is equal to $Mgl b/a$ at turning angles and $\dot \phi$ will be equal to $0$ at one of the turning angles and $2 Mgl /(I_x a)$ at the other one.
$\dot \phi=0$ at $\theta=\theta_{min}$, and $\dot \phi$ is equal to $2 Mgl /(I_x a)$ at $\theta=\theta_{max}$ when $Mgl>0$ which can be understood from equation \eqref{ddottheta}, i.e. if $\dot \phi=0$ then $\ddot \theta>0$ and $\theta$ increases.
Therefore, $\dot \phi$ will take values between $0$ and $2 Mgl /(I_x a)$ (see appendix 1), and it will change according to the equation \eqref{phidot}.
Change of $\dot \psi$ can be found by using equation \eqref{psidot}.

The simplest way of observing this type of motion is setting a heavy symmetric top, spinning with $\dot \psi_0$, into motion with an inclination angle $\theta_0$ without nutation and precession angular velocities $\dot \theta_0=0$ and $\dot \phi_0=0$ \cite{Goldstein}.
In this case, $\theta_0$ is the minimum angle because of the presence of the gravitational force.
At the beginning, $b$ will be equal to $a \cos \theta_0$ or $\theta_0=\arccos (b/a)$ since $\dot \phi_0=0$.
With these initial conditions, according to equation \eqref{eprime}, $E'$ becomes equal to $Mgl b/a$.

A case with initial values $\theta_0=0.524 \,rad$, $\dot \theta_0=0$, $\dot \phi_0=0$ and $\dot \psi_0=250 \,rad \,s^{-1}$ will be considered as an example.  
In this case, $a=62.7 \,rad \,s^{-1}$, $b=54.3 \,rad \,s^{-1}$ and $E'=Mglb/a=0.0589 \,J$.  

\begin{figure}[!h]
	\begin{center}
		\subfigure[]{
			\includegraphics[width=4.2cm]{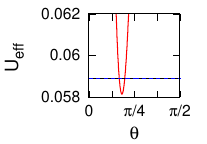}
		}
		\subfigure[]{
			\includegraphics[width=4.2cm]{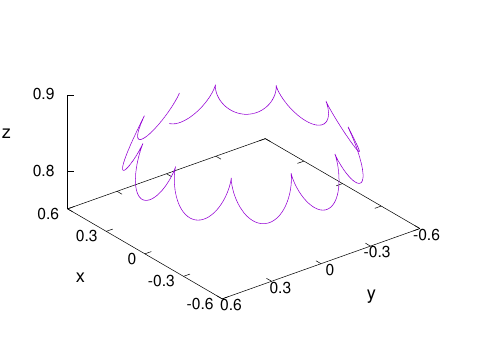}
		}
		\subfigure[]{
			\includegraphics[width=4.2cm]{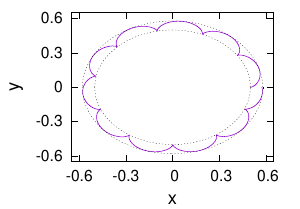}
		}
		\caption{(a) $U_{eff}$ (red curve), $E'$ (dashed blue line) and $Mgl b/a$ (black line).
		(b) Shapes for the locus on the unit sphere for the motion with cusps. 
		(c) Projection of shapes for the locus on $x'y'$-plane.
		Initial values are $\theta_0=0.524 \,rad$, $\dot \theta_0=0$, $\dot \phi_0=0$ and $\dot \psi_0=250 \,rad \,s^{-1}$, and constants are $a=62.7 \,rad \,s^{-1}$, $b=54.3 \,rad \,s^{-1}$ and $E'=Mglb/a=0.0589 \,J$.
		The animated plot is available at \href{https://youtu.be/ZkEImVMsnzo}{https://youtu.be/ZkEImVMsnzo}.
		}
		\label{fig:uefftt_1}
	\end{center}
\end{figure}

In figure \ref{fig:uefftt_1}(a), the effective potential, $E'$ and $Mgl b/a$ are shown.
Turning angles of $\theta$ can be obtained as $\theta_{min}=0.524 \,rad$ and $\theta_{max}=0.618 \,rad$ by using $E'=U_{eff}(\theta_{ext})$ or $f(u)$ and taking into account values of constants.

The results of numerical solutions for variables can be found in figure \ref{fig:ttdfd_1} in appendix 4.
These graphs are drawn for one nutation period, and these changes repeat themselves at each nutation period.
$\theta$ changes between $\theta_{min}$ and $\theta_{max}$.
As it is expected, $\dot \phi$ starts from $0$ and increases till $2 Mgl/(I_x a)=9.51 \,rad \,s^{-1}$, which can be seen in figure \ref{fig:ttdfd_1}(e).
The changes in $\dot \psi$ and $\dot \phi$ can be understood by using $a$ and $b$, and their changes obey the conservation of angular momenta.
In general, $\dot \psi$ and $\dot \phi$ can show different behaviour (see appendix 1 and 2), but in this case, the situation is simple, and $\dot \phi$ increases while $\dot \psi$ decreases, see figures \ref{fig:ttdfd_1}(e) and \ref{fig:ttdfd_1}(f).
For the second half of the nutation period, the changes in $\dot \phi$ and $\dot \psi$ is in the reverse order of the first half of the nutation period.

The three-dimensional figures for shapes for the locus are drawn by using the results of the first numerical solution, figure \ref{fig:uefftt_1}(b).
Its projection can be seen in figure \ref{fig:uefftt_1}(c).
It can be seen that there are cusps due to zeros of $\dot \phi$ at the minimum of $\theta$.

At the beginning, we have taken that $\dot \theta_0=0$ and $\dot \phi_0=0$, but motion with cusps is also possible if initial $\dot \theta_0$ and $\dot \phi_0$ are not equal to $0$ provided that $E'=Mgl b/a$ and $|a|>|b|$.

\subsubsection{Wavy precession}
\label{wp}

Now, we will consider a case satisfying $E'<Mglb/a$ and $|a|>|b|$.
$\dot \phi$ can be equal to $0$ when $U_{eff}=Mglb/a$ and $|a|>|b|$, and $U_{eff}$ can not be equal to $Mglb/a$ when $E'<Mglb/a$.
Then, in this case, $\dot \phi$ can not be equal to $0$ and change sign.
Therefore, $\dot \phi$ always has the same sign, and the spinning heavy symmetric top precesses in one direction with changing $\dot \phi$.
In such cases, there is more than one nutation in one precession period, and a wavy pattern is seen.

\begin{figure}[!h]
	\begin{center}
		\subfigure[]{
			\includegraphics[width=4.2cm]{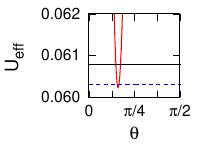}
		}
		\subfigure[]{
			\includegraphics[width=4.2cm]{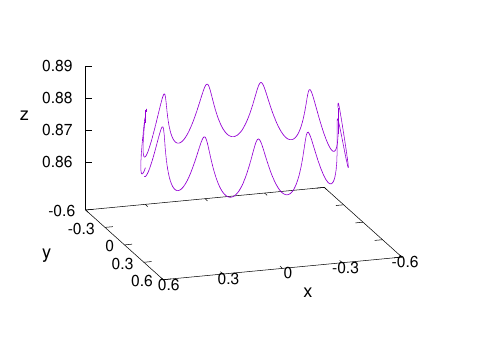}
		}
		\subfigure[]{
			\includegraphics[width=4.2cm]{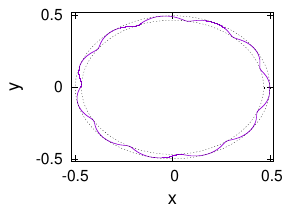}
		}
		\caption{(a) $U_{eff}$ (red curve), $E'$ (dashed blue line) and $Mgl b/a$ (black line).
		(b) Shapes for the locus on the unit sphere for wavy precession. 
		(c) Projection of shapes for the locus on $x'y'$-plane.
		Initial values are $\theta_0=0.524 \,rad$, $\dot \theta_0=0$, $\dot \phi_0=7.00 \,rad \,s^{-1}$ and $\dot \psi_0=250 \,rad \,s^{-1}$, and constants are $a=64.2 \,rad \,s^{-1}$, $b=57.4 \,rad \,s^{-1}$, $E'=0.0603 \,J$ and $Mgl b/a=0.0608 \,J$.
		The animated plot is available at \href{https://youtu.be/QFh_lXjwr-g}{https://youtu.be/QFh\_lXjwr-g}.
		}
		\label{fig:uefftt_2}
	\end{center}
\end{figure}

To obtain such a case, one can choose initial values as $\theta_0=0.524 \,rad$, $\dot \theta_0=0$, $\dot \phi_0=7.00 \,rad \,s^{-1}$ and $\dot \psi_0=250 \,rad \,s^{-1}$.
With these initial values, constants can be found as $a=64.2 \,rad \,s^{-1}$, $b=57.4 \,rad \,s^{-1}$, $E'=0.0603 \,J$ and $Mgl b/a=0.0608 \,J$.
The turning angles of $\theta$ are $\theta_{max}=0.524 \,rad$ and $\theta_{min}=0.488 \,rad$.
$U_{eff}$, $E'$ and $Mglb/a$ can be seen in figure \ref{fig:uefftt_2}(a).

The results of numerical solutions are shown in figure \ref{fig:ttdfd_2} in appendix 4. 
It can be seen that $\theta$ changes between turning angles, and $\dot \theta$ takes negative and positive values in accordance with equation \eqref{eprime}.
$\dot \phi$ never becomes zero or changes sign and it is always greater than $0$ and smaller than $2 Mgl/(I_xa)$.
The relationship between conserved angular momenta and angular velocities $\dot \phi$ and $\dot \psi$ are similar to the motion with cusps.

The three-dimensional plot and its projection are available in figures \ref{fig:uefftt_2}(b) and \ref{fig:uefftt_2}(c), respectively.
The wavy structure of the precession is visible in the three-dimensional figure and its projection.
For wavy precession, the difference between $E'$ and the minimum of $U_{eff}$ can be small with respect to other types of motion, and then the oscillations in $\theta$ also become small.

\subsubsection{Looping motion}
\label{loopingmotion}

To obtain looping motion, we will consider a case satisfying $E'>Mglb/a$ and $|a|>|b|$.
In this case, $U_{eff}$ becomes equal to $Mglb/a$ at two different angles which are between turning angles.
And, in one of these angles, $\dot \phi$ becomes zero, and possible interval of $\theta$ includes angles from both sides of that angle since $E'>Mglb/a$, and $\dot \phi$ is negative in one side of that angle and it is positive at the other side (see appendix 1).
Then, the spinning heavy symmetric top makes a looping motion, and there is more than one nutation in one precession period.

\begin{figure}[!h]
	\begin{center}
		\subfigure[]{
			\includegraphics[width=4.2cm]{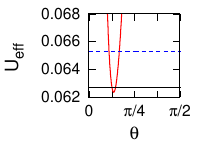}
		}
		\subfigure[]{
			\includegraphics[width=4.2cm]{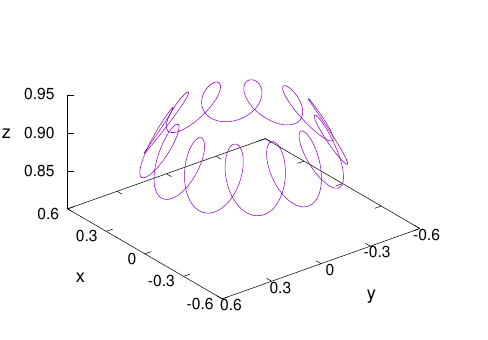}
		}
		\subfigure[]{
			\includegraphics[width=4.2cm]{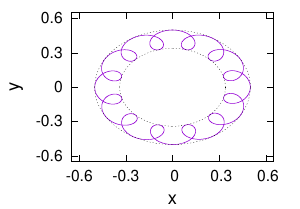}
		}
		\caption{(a) $U_{eff}$ (red curve), $E'$ (dashed blue line) and $Mgl b/a$ (black line).
		(b) Shapes for the locus on the unit sphere for looping motion. 
		(c) Projection of shapes for the locus on $x'y'$-plane.
		Initial values are $\theta_0=0.524 \,rad$, $\dot \theta_0=0$, $\dot \phi_0=15.0 \,rad \,s^{-1}$ and $\dot \psi_0=250 \,rad \,s^{-1}$, and constants are $a=66.0 \,rad \,s^{-1}$, $b=60.9 \,rad \,s^{-1}$, $E'=0.0653 \,J$ and $Mgl b/a=0.0627 \,J$.
		The animated plot is available at \href{https://youtu.be/Y1t0el21L3g}{https://youtu.be/Y1t0el21L3g}.
		}
		\label{fig:uefftt_3}
	\end{center}
\end{figure}

To obtain such a case, one can take following initial values: $\theta_0=0.524 \,rad$, $\dot \theta_0=0$, $\dot \phi_0=15.0 \,rad \,s^{-1}$ and $\dot \psi_0=250 \,rad \,s^{-1}$.
With these initial values, the constants can be found as $a=66.0 \,rad \,s^{-1}$, $b=60.9 \,rad \,s^{-1}$, $E'=0.0653 \,J$ and $Mgl b/a=0.0627 \,J$.
The turning angles of $\theta$ are $\theta_{max}=0.524 \,rad$ and $\theta_{min}=0.347 \,rad$.
Graph of $U_{eff}$ together with $E'$ and $Mgl b/a$ can be seen in figure \ref{fig:uefftt_3}(a).

The results of numerical solutions can be seen in figure \ref{fig:ttdfd_3} in appendix 4.
$\dot \phi$ becomes zero when $b$ is equal to $a \cos \theta$, and with these initial values, it occurs at $\theta=\arccos (b/a)=0.3957 \,rad$.
Numerical solutions show that when $\theta<0.3957 \,rad$, $\dot \phi$ is negative; and it is positive when $\theta>0.3957 \,rad$ as expected (see appendix 1).

The three-dimensional plot and its projection can be seen in figure \ref{fig:uefftt_3}(b) and \ref{fig:uefftt_3}(c).
The looping structure can be seen in these figures.

As mentioned, this type of motion is observed when $E'>Mglb/a$ and $|a|>|b|$.
As $E'$ gets closer to $Mglb/a$, loops become smaller; and as $E'$ takes bigger values, bigger loops are observed, and they start to overlap.
The overall precession has the same sign as $a$ \cite{Leimanis}, and it is independent of the greatness of the difference $E'-Mglb/a$.

\subsection{Different types of motion when $|b|>|a|$}
\label{bgraterthana}

In this section, different types of motions will be studied when $|b|>|a|$. 
In this case, $\dot \phi$ never becomes zero and never changes sign (see appendix 1).
Therefore, shapes for the locus do not show similar properties to cases satisfying $|a|>|b|$.
However, there are still interesting properties of motion, and one of them is the spin reversal.
We will also consider a case satisfying $E'=Mglb/a$, but the relation between $E'$ and $Mglb/a$ does not affect shapes for the locus similar to previous cases since a small difference in $E'$ does not cause a distinguishable change in them.
We should note that when $|b|>|a|$, $Mglb/a$ can be smaller than the minimum of $U_{eff}$ \cite{Tanriverdi_ueff}, which, in general, happens for grater values of $|b|$ and $|a|$.
When $|b|>|a|$, the precession direction has always the same sign as $b$ or $\dot \phi$ (see appendix 1).

\subsubsection{Regular precession when $|b|>|a|$}

We studied the details of the regular precession above.
Here, we will give an example of the regular precession when $|b|>|a|$.

If we take initial values as follows $\theta_0=0.524 \,rad$, $\dot \theta_0=0$, $\dot \phi_0=91.7 \,rad \,s^{-1}$ and $\dot \psi_0=250 \,rad \,s^{-1}$, the constants become $a=82.6 \,rad \,s^{-1} $, $b=94.5 \,rad \,s^{-1}$, $E'=0.299 \,J$ and $Mglb/a=0.0778 \,J$.
$\theta_0$, $\dot \theta_0$ and $\dot \psi_0$ are the same as the regular precession considered in section \ref{regpr1} and $\dot \phi_0$ is the second root obtained from equation \eqref{regular12}.
In this case, $Mglb/a$ is smaller than the minimum of $U_{eff}$.

\begin{figure}[!h]
	\begin{center}
		\subfigure[]{
			\includegraphics[width=4.2cm]{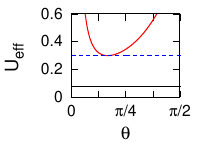}
		}
		\subfigure[]{
			\includegraphics[width=4.2cm]{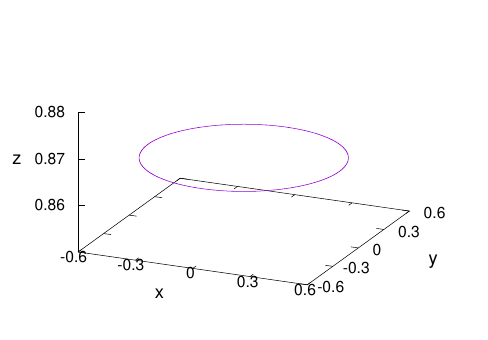}
		}
		\caption{(a) $U_{eff}$ (red curve), $E'$ (dashed blue line) and $Mgl b/a$ (black line).
		(b) Shapes for the locus on the unit sphere for the regular precession when $a<b$. 
		Initial values are $\theta_0=0.524 \,rad$, $\dot \theta_0=0$, $\dot \phi_0=91.7 \,rad \,s^{-1}$ and $\dot \psi_0=250 \,rad \,s^{-1}$, and constants are $a=82.6 \,rad \,s^{-1}$, $b=94.5 \,rad \,s^{-1}$, $E'=0.299 \,J$ and $Mglb/a=0.0778 \,J$.
		The animated plot is available at \href{https://youtu.be/C7588R3XI3U}{https://youtu.be/C7588R3XI3U}.
		}
		\label{fig:uefftt_7a2}
	\end{center}
\end{figure}

In figure \ref{fig:uefftt_7a2}(a), one can see the graph of $U_{eff}$ together with $E'$ and $Mgl b/a$,
and one can see that $E'$ is equal to the minimum of $U_{eff}$.
From figures \ref{fig:uefftt_7a1}(a) and \ref{fig:uefftt_7a2}(a), it can be seen that form and the minimum value of $U_{eff}$ are changed since $\dot \phi$ is different, but the position of the minimum is not changed.
This shows that both regular precessions take place at the same inclination angle as expected.

Results of the numerical integration of angular accelerations can be seen in figure \ref{fig:ttdfd_7a2} in appendix 4.
The three-dimensional plot of shapes for the locus can be seen in figure \ref{fig:uefftt_7a2}(b).
It can be seen that the top precesses regularly.

From equation \eqref{phidot} or appendix 1, one can say that the precession direction has the same sign as $b$ when $b>a$, and we have already mentioned that it has the same sign as $a$ when $a>b$.
In the given examples, both $a$ and $b$ are positive, then both roots of $\dot \phi$ become positive.
When $a$ and $b$ have different signs, one of the roots of equation \eqref{regular12} has a different sign, and then one of the regular precession takes place in one direction, and the other one takes place in the other direction \cite{ArnoldMaunder}.
The angle restrictions for these cases are explained in section \ref{regpr1}.

\subsubsection{Motion with the same precession angular velocity at both extrema}
\label{mwepavam}

To obtain motion with the same precession angular velocity at both extrema, we will consider a case satisfying $E'=Mglb/a$ when $|b|>|a|$.
Since $|b/a|>1>|\cos\theta|$, in this case, $\dot \phi=(b- a\cos \theta)/\sin^2 \theta$ cannot be equal to $0$.
Then, from equation \eqref{mglboadotphi}, $\dot \phi$ can be obtained at both extrema of $\theta$ as
\begin{equation}
	\dot \phi (\theta=\theta_{ext})=\frac{2 Mgl }{I_x a}.
\end{equation}
Then, a motion with the same precessional angular velocity at extrema of $\theta$ will be observed when $|b|>|a|$ and $E'=Mgl b/a$.

If initial values are chosen as $\theta_0=0.579 \,rad$, $\dot \theta=0$, $\dot \phi=25.9 \,rad \,s^{-1}$ and $\dot \psi= 70.2 \,rad \,s^{-1}$, this type of motion can be obtained.
Initial value of $\dot \phi$ is equal to $2 Mgl/(I_x a)$.
In this case, $a=23.0 \,rad \,s^{-1}$, $b=27.0 \,rad \,s^{-1}$ and $E'=Mgl b/a=0.0798 \,J$.
For this case turning angles become $\theta_{min}=0.579 \,rad$ and $\theta_{max}=1.52 \,rad$.

\begin{figure}[!h]
	\begin{center}
		\subfigure[]{
			\includegraphics[width=4.2cm]{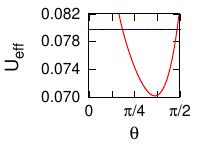}
		}
		\subfigure[]{
			\includegraphics[width=4.2cm]{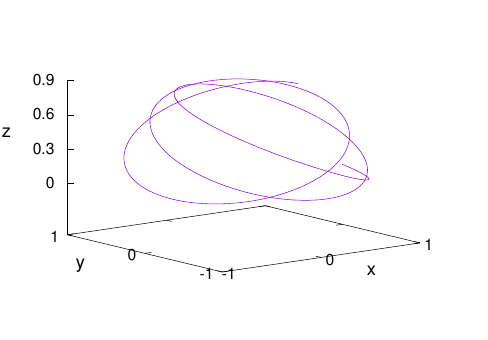}
		}
		\subfigure[]{
			\includegraphics[width=4.2cm]{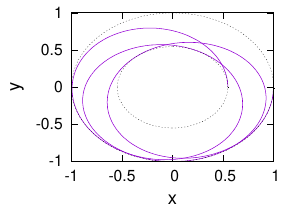}
		}
		\caption{(a) $U_{eff}$ (red curve), $E'$ (dashed blue line) and $Mgl b/a$ (black line).
		(b) Shapes for the locus on the unit sphere for motion with the same precession angular velocity at both extrema. 
		(c) Projection of shapes for the locus on $x'y'$-plane.
		Initial values are $\theta_0=0.579 \,rad$, $\dot \theta_0=0$, $\dot \phi_0=25.9 \,rad \,s^{-1}$ and $\dot \psi_0=70.2 \,rad \,s^{-1}$, and constants are $a=23.0 \,rad \,s^{-1}$, $b=27.0 \,rad \,s^{-1} $ and $E'=Mglb/a=0.0798 \,J$.
		The animated plot is available at \href{https://youtu.be/vL2FPV6SaCU}{https://youtu.be/vL2FPV6SaCU}.
		}
		\label{fig:uefftt_4}
	\end{center}
\end{figure}

$U_{eff}$, $E'$ and $Mglb/a$ can be seen in figure \ref{fig:uefftt_4}(a), and it can be seen that $E'$ is equal to $Mglb/a$.

Results of numerical solutions can be seen in figure \ref{fig:ttdfd_4} in appendix 4. 
Unlike cases satisfying $|a|>|b|$, in which $\dot \phi$ either increases or decreases during the half-nutation period, in this case $\dot \phi$ decreases at the beginning and then increases to its starting value during the half-nutation period.

Shapes for the locus for a few precession period and their projection can be seen in figures \ref{fig:uefftt_4}(b) and \ref{fig:uefftt_4}(c).
At these figures, it can be seen that there is nearly one nutation at each precession.
These shapes for the locus also give an example of the general structure of motion when $|b|>|a|$. 

\subsubsection{Other possible motions when $|b|>|a|$}
\label{opm}

As mentioned previously, when $|b|>|a|$, the difference in shapes for the locus is not easily distinguishable similar to cases when $|a|>|b|$.
However, the differences should be noted, and by using this example, we will mention them and give an example of the spin reversal.

\begin{figure}[!h]
	\begin{center}
		\subfigure[]{
			\includegraphics[width=4.2cm]{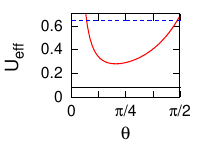}
		}
		\subfigure[]{
			\includegraphics[width=4.2cm]{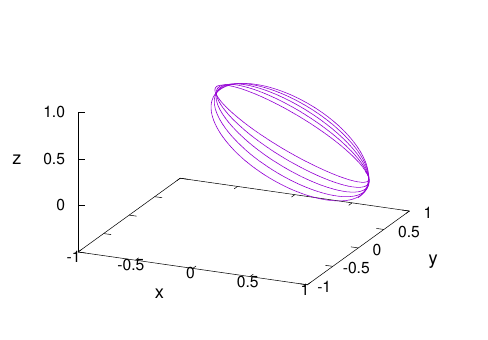}
		}
		\subfigure[]{
			\includegraphics[width=4.2cm]{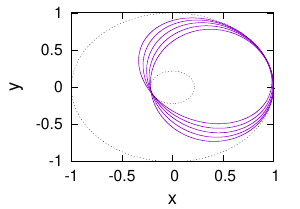}
		}
		\caption{(a) $U_{eff}$ (red curve), $E'$ (dashed blue line) and $Mgl b/a$ (black line).
		(b) Shapes for the locus on the unit sphere for spin reversing motion when $b>a$.
		(c) Projection of shapes for the locus on $x'y'$-plane. 
		Initial values are $\theta_0=1.53 \,rad$, $\dot \theta_0=0$, $\dot \phi_0=75.0 \,rad \,s^{-1}$ and $\dot \psi_0=250 \,rad \,s^{-1}$, and constants are $a=63.5 \,rad \,s^{-1}$, $b=77.5 \,rad \,s^{-1}$, $E'=0.643 \,J$ and $Mgl b/a=0.0830 \,J$.
		The animated plot is available at \href{https://youtu.be/YpoJEgdmnuw}{https://youtu.be/YpoJEgdmnuw}.
		}
		\label{fig:uefftt_sr_b1}
	\end{center}
\end{figure}

If we take initial values as follows $\theta_0=1.53 \,rad$, $\dot \theta_0=0$, $\dot \phi_0=75.0 \,rad \,s^{-1}$ and $\dot \psi_0=250 \,rad \,s^{-1}$, then the constants can be obtained as $a=63.5 \,rad \,s^{-1}$, $b=77.5 \,rad \,s^{-1}$, $E'=0.643 \,J$ and $Mgl b/a=0.0830 \,J$.
With these initial values, turning angles become $\theta_{max}=1.53 \,rad$ and $\theta_{min}=0.220 \,rad$.
In this case, $Mgl b/a$ is smaller than the minimum of $U_{eff}$ similar to the regular precession when $|b|>|a|$, which can be seen in figure \ref{fig:uefftt_sr_b1}(a) together with $E'$.

Results of numerical solutions can be seen in figure \ref{fig:ttdfd_sr_b1} in appendix 4.
It can be seen from figure \ref{fig:ttdfd_sr_b1}(f) that spin direction is changing, and by using equation \eqref{rootdotpsi}, one can calculate the angle where spin reversal takes place at $\theta=0.250\, rad$ and it is consistent with numerical calculations.
In figure \ref{fig:ttdfd_sr_b1}(e), it can be seen that $\dot \phi$ shows a significant increase at the end of the half-nutation period.
This increase takes place together with the change in the spin direction.
More detailed analyses of spin reversal when $|b|>|a|$ and experimental results can be found in previous works \cite{Cross2013, Tanriverdi2019, Tanriverdi2020b} (see also appendix 2).

Shapes for the locus and their projection can be seen in figures \ref{fig:uefftt_sr_b1}(b) and \ref{fig:uefftt_sr_b1}(c) for a few precession period, respectively.
It can also be seen that there is a small advance in the precession at each nutation period.

In some other cases when $|b|>|a|$, there can be only an increase or decrease in $\dot \phi$ during the half-nutation period.
Spin reversal may or may not be seen in similar cases depending on $E'$.

If $|b|$ is more grater than $|a|$ compared to this case, then the difference between the nutation period and precession period will be smaller, and loops will be a little advanced with respect to the overall precession direction.
If $|b|$ is very close to $|a|$, the shapes for the locus will be more similar to motion with equal precession angular velocity at both extrema.

\subsection{Motion with negative $Mgl$}

\begin{figure}[!h]
	\begin{center}
		\subfigure[]{
			\includegraphics[width=4.2cm]{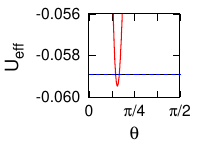}
		}
		\subfigure[]{
			\includegraphics[width=4.2cm]{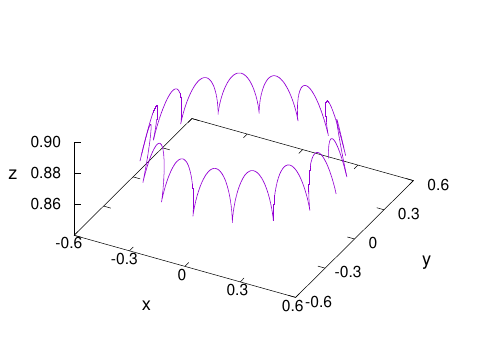}
		}
		\subfigure[]{
			\includegraphics[width=4.2cm]{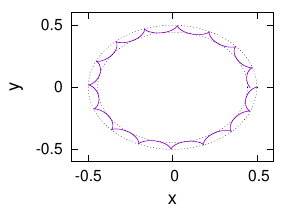}
		}
		\caption{(a) $U_{eff}$ (red curve), $E'$ (dashed blue line) and $Mgl b/a$ (black line).
		(b) Shapes for the locus on the unit sphere for motion with negative $Mgl$. 
		(c) Projection of shapes for the locus on $x'y'$-plane.
		Initial values are $\theta_0=0.524 \,rad$, $\dot \theta_0=0$, $\dot \phi_0=0$ and $\dot \psi_0=250 \,rad \,s^{-1}$, and constants are $a=62.7 \,rad \,s^{-1}$, $b=54.3 \,rad \,s^{-1}$, $E'=-0.0589 \,J$ and $Mgl b/a=-0.0589 \,J$.
		Turning angles are $\theta_{max}=0.524 \,rad$ and $\theta_{min}=0.461 \,rad$.
		The animated plot is available at \href{https://youtu.be/sqMJTymRPPM}{https://youtu.be/sqMJTymRPPM}.
		}
		\label{fig:uefftt_mmgl}
	\end{center}
\end{figure}

When $Mgl$ is negative, changes in the motion are seen since the symmetric top is forced to move upwards differently from the positive $Mgl$ case \cite{KleinSommerfeld, Tanriverdi2020, Moralesea2016}.
This is not possible for an ordinary symmetric top. 
However, it is possible for gyroscopes and some special symmetric tops with extra pieces.
We will consider the initial values used for the motion with cusps as an example to motion with negative $Mgl$, and $Mgl$ will be taken as $-0.068 \,J$.
These initial values can be found in the explanations of figure \ref{fig:uefftt_mmgl}.

Effective potential together with $E'$ and $Mglb/a$ for this case can be seen in figure \ref{fig:uefftt_mmgl}(a).
It can be seen that the minimum of effective potential is negative, which is different from the motion with cusps studied in section \ref{mwc}.
The position of $U_{eff_{min}}$ and form of $U_{eff}$ are also different.
More detailed analysis of effective potential when $Mgl$ is negative can be found in previous work \cite{Tanriverdi_ueff}.

Results of numerical solutions can be seen in figure \ref{fig:ttdfd_mmgl} in appendix 4.
It can be seen in figure \ref{fig:ttdfd_mmgl}(a) that $\dot \theta$ is negative at the beginning and $\theta$ decreases after the settlement of the motion differently from the motion with cusps.
This happens since negative $Mgl$ applies torque to lift the symmetric top or gyroscope.

In this case, $\dot \phi$ takes negative values, and symmetric top precesses in the reverse direction with respect to precession in the motion with cusps.

In figures \ref{fig:uefftt_mmgl}(b) and \ref{fig:uefftt_mmgl}(c), shapes for the locus and their projection can be seen, respectively.
From these figures, it can be seen that an inverted motion with cusps is observed.

In other cases with negative $Mgl$, other inverted shapes for the locus can be obtained.
In the given example, $E'=Mglb/a$ and $|a|>|b|$ relations are satisfied, and an inverted motion with cusps is observed.
From equations \eqref{ueffmglboa} and \eqref{mglboadotphi}, one can say that the sign of $Mgl$ does not affect these relations, and $\dot \phi$ becomes equal to zero at one of the inclination angles making $U_{eff}=Mglb/a$.
When $Mgl$ is negative, the general structure of effective potential does not change, but the position and the magnitude of the minimum of $U_{eff}$ change \cite{Tanriverdi_ueff}.
From equations \eqref{ddottheta}, \eqref{ddotphi} and \eqref{ddotpsi}, one can see that $Mgl$ is only present in the equation related to angular acceleration $\ddot \theta$.
These show that negative $Mgl$ directly affects change in inclination angle which results in inverted motions.
$\dot \phi$ and $\dot \psi$ are affected indirectly, and the precession direction can be in the reverse direction, and spin angular velocity can increase instead of decreasing similar to the given example.
Since the condition for $\dot \phi=0$ remains the same and motion type mainly determined by zeros of $\dot \phi$ and $\dot \theta$, the conditions which are used to determine motion type for the cases when $Mgl>0$ can still be used when $Mgl<0$.

In some cases, inverted shapes can easily be noticed, e.g., the motion with cusps or looping motion.
However, for cases like wavy precession or cases satisfying $|b|>|a|$, one may not understand the inverted structure by looking at shapes for the locus.

\section{Conclusion}
\label{rd}

A symmetric top or gyroscope can start its motion with different initial values and it can make different motions, or the same motion.
One can not understand the motion type by only looking at the initial values of dynamical variables, and consideration of constants of motion is necessary.
On the other hand, only considering constants of motion can cause misinterpretation.
While studying the regular precession, we have seen that different possibilities are shadowed because of the usage of constants.
Therefore, for complicated systems like the motion of a heavy symmetric top, it can be necessary to consider both dynamical variables and constants.

One of the interesting dynamical changes is the spin reversal.
And, in this study, we have given an example of this while studying different types of motion.
In appendix 2, we have studied the changes in $\dot \psi$, and one can say that spin reversal can only take place when $|b|>|a|$ and possible $\theta$ interval includes the root given by equation \eqref{rootdotpsi}.
$\dot \phi$ can also show an interesting change, and "motion with the same precession angular momenta at both extrema" gives an example of previously unnoticed change.
Probably, this motion and the spin reversal is not noticed till this time because of considering only constants but not dynamical variables.
On the other hand, the mentioned unnoticed change in $\dot \phi$ is obtained by considering a relation equivalent to Routh's one, who studied the motion in terms of geometric constants.
In appendix 1, we have studied the general structure of the changes in $\dot \phi$.
Form studies in these appendixes, it can be seen that one of the main differences between $\dot \phi$ and $\dot \psi$ is related with the sign change condition: $\dot \phi$ can change sign when $|a|>|b|$ and $\dot \psi$ can change sign when $|b|>|a|$.

The general classification of motion type is mainly determined by nutation and precession angular velocities. 
This classification depends on the shapes for the locus, and by using constants of motion one can determine the type of shapes for the locus.
We have seen that this determination can be done by considering $a$, $b$, $E'$, $Mglb/a$ and $U_{eff_{min}}$ when $|b|\ne|a|$. 

The necessary condition for the regular precession can be considered as the equivalence of $E'$ and $U_{eff_{min}}$ when $|b|\ne|a|$.
If $E'>U_{eff_{min}}$, there can be various types of motion.
We have seen that there are three possible motions when $|a|>|b|$ and $E'>U_{eff_{min}}$: motion with cusps ($E'=Mglb/a$), wavy precession ($E'<Mglb/a$) and looping motion ($E'>Mglb/a$).
These three different types of motion are always possible since $Mglb/a$ is always greater than the minimum of $U_{eff}$ \cite{Tanriverdi_ueff}.
When $|b|>|a|$, sign change of $\dot \phi$ is not possible and a small change in $E'$ does not result in a distinguishable change in shapes for the locus similar to cases satisfying $|a|>|b|$.
Though the difference in shapes for the locus is less, besides the regular precession, we studied two different cases when $|b|>|a|$ to be able to mention different possibilities.
And, numerical solutions have shown that one nutation period is longer than one precession period when $|b|>|a|$ differently from $|a|>|b|$ cases.
Another interesting change is seen when $Mgl<0$, and we have seen that usage of employed constants can also be used to determine inverted motions.

We should note that all initial values have resulted in positive values of $a$ and $b$.
When one or both of these two constants are negative, similar shapes for the locus with a different spin and/or precession direction can be obtained if relations between constants are the same.
When one of these two constants is negative, $Mglb/a$ becomes negative, and it can be necessary to choose initial values giving $E'$ as negative.
We should also note that all relations between $Mglb/a$ and $E'$ can not be satisfied always: $Mglb/a$ can be smaller than $U_{eff_{min}}$ when $|b|>|a|$ \cite{Tanriverdi_ueff}, and in such cases, it is impossible to satisfy relations $E'=Mglb/a$ and $E'<Mglb/a$.  

We have mentioned that four parameters have been used to classify motion types in Routh's work \cite{Routh}.
We have given a detailed comparison of parameters and motion types in appendix 3.
We have seen that usage of parameters is equivalent and figures are consistent for motion with cusps and looping motion, where $|a|>|b|$.
We have also seen that two cases satisfying $|b|=|a|$ are considered at there, and conditions and figures are consistent in general properties with previous work \cite{Tanriverdi2020}.
The situation when $|b|>|a|$ is different.
We already mentioned that $E'$ does not affect shapes for the locus in a distinguishable way, and then we did not consider all possibilities though Routh has considered.
For cases satisfying $|b|>|a|$, with some small differences, the conditions and figures are similar.
One of Routh's parameters, $[OQ]$, is inconsistent with the ones considered in this work.
And, the examples in appendix 3 have shown that Routh's parameter $[OQ]$ should be replaced with $[a]$ to determine the motion type correctly. 

\section{Acknowledgment}

Thanks to anonymous referees for their helpful comments.

\section{Appendix 1}
\label{ap1}

\begin{figure}[!h]
	\begin{center}
		\subfigure[$|a|>|b|$]{
			\includegraphics[width=5.2cm]{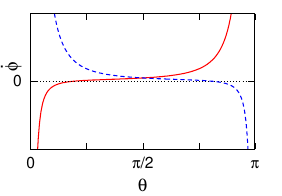}
		}
		\subfigure[$|b|>|a|$]{
			\includegraphics[width=5.2cm]{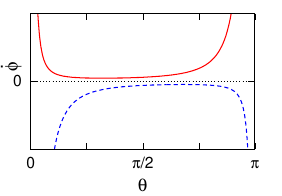}
		}
		\caption{Change of $\dot \phi$ with respect to $\theta$ when (a) $|a|>|b|$ and (b) $|b|>|a|$.
	                (a) Continuous (red) curve shows $\dot \phi$ when $a>b>0$, and dashed (blue) curve shows $\dot \phi$ when $a<0$ and $b>0$. 
	                (b) Continuous (red) curve shows $\dot \phi$ when $b>a>0$, and dashed (blue) curve shows $\dot \phi$ when $a>0$ and $b<0$. 
		}
		\label{fig:cdp}
	\end{center}
\end{figure}

In this part, we will study the change of $\dot \phi$ by using $a$ and $b$ when their magnitudes are not equal.
In section \ref{eom}, we have seen that $\dot \phi$ can be written in terms of $a$ and $b$ as $\dot \phi(\theta)=(b-a \cos \theta)/\sin^2 \theta$, where $\theta \in [0,\pi]$.
It can be better to take the right-hand side of this equation in $a$ parentheses, and then one can write
\begin{equation}
	\dot \phi(\theta)=a\left( \frac{b/a -  \cos \theta}{\sin^2 \theta} \right).
	\label{phidota}
\end{equation}
It can be seen that there is not any root when $|b|>|a|$ and there is always a root when $|b|<|a|$ since $-1 < \cos \theta < 1$ for possible $\theta$ values, and that root can be found as $\theta_r=\arccos(b/a)$.
One can easily say that the root is between $0$ and $\pi/2$ when $1>b/a>0$, and it is between $\pi/2$ and $\pi$ when $-1<b/a<0$.  
To get the root at $\pi/2$, $b$ should be equal to $0$.
When $|a|>|b|$, $\dot \phi$ gets negative and positive values at different regions of the domain of $\theta$ since $\cos \theta>b/a$ when $\theta<\theta_r$ and $\cos \theta<b/a$ when $\theta>\theta_r$, and $\dot \phi$ goes to $-sgn(a)\infty$ as $\theta$ goes to $0$, and $\dot \phi$ goes to $sgn(a)\infty$ as $\theta$ goes to $\pi$.
When $|b|>|a|$, $\dot \phi$ has always the same sign as $b$ since $|b|>|a \cos \theta|$, and it goes to $sgn(b)\infty$ as $\theta$ goes to $0$ or $\pi$.
One can see four different possibilities in figure \ref{fig:cdp}.

The derivative of $\dot \phi$ with respect to $\theta$ can be written as
\begin{equation}
	\frac{d \dot \phi(\theta)}{d \theta}=\frac{a}{\sin^3 \theta} \left( \cos^2 \theta -2 \frac{b}{a} \cos \theta+1\right).  
	\label{derphidot}
\end{equation}
The root of this equation, corresponding to the turning point, can be found by using the terms in the parentheses as
\begin{equation}
	\theta_{tp}=\arccos \left(\frac{b}{a}\pm \sqrt{\frac{b^2}{a^2}-1} \right).    
	\label{rodph}
\end{equation}
From the terms inside the square root, one can say that there is not any turning point when $|a|>|b|$, and there is always a turning point when $|b|>|a|$.
The turning point is between $0$ and $\pi/2$ when $b/a>1$, and it is between $\pi/2$ and $\pi$ when $b/a<-1$. 
Two possible cases can be seen in figure \ref{fig:cdp}(b).

During the motion of the symmetric top, $\dot \phi$, in general, does not show all of the changes seen in one of the curves shown in figure \ref{fig:cdp}.
It takes values according to the possible $\theta$ values.

By considering these evaluations, one can say that $\dot \phi$ can either increase or decrease during the half nutation period, and it can be $0$ if possible $\theta$ values include $\arccos(b/a)$ when $|a|>|b|$.
$\dot \phi$ can either increase or decrease, or can both increase and decrease during the half nutation period, and it can not be $0$ when $|b|>|a|$.

\section{Appendix 2}
\label{ap2}

In this part, we will study the change of $\dot \psi$ when magnitudes of $a$ and $b$ are not equal.
In section \ref{eom}, $\dot \psi$ is given as $\dot \psi(\theta)=(I_x/I_z)a-(b - a \cos \theta) \cos\theta / \sin^2 \theta$, where $\theta \in [0,\pi]$.
For a symmetric top, one can write $I_x/I_z=1/2+\epsilon$, where $\epsilon>0$ \cite{Tanriverdicorridengum}, and then, $\dot \psi(\theta)$ can be written as
\begin{equation}
	\dot \psi(\theta)=a\left( \frac{\cos^2 \theta-\frac{2b}{a} \cos \theta+1}{ 2 \sin^2 \theta}+\epsilon \right).
	\label{psidot2}
\end{equation}
Understanding the changes in $\dot \psi$ is easier in this form.

\begin{figure}[!h]
        \begin{center}
        \subfigure[$|a|>|b|$]{
               \includegraphics[width=5.2cm]{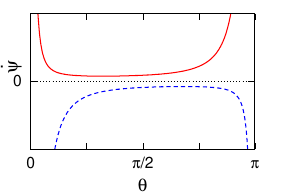}
        }
        \subfigure[$|b|>|a|$]{
               \includegraphics[width=5.2cm]{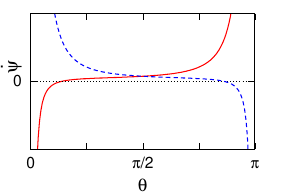}
        }
        \caption{Change of $\dot \psi$ with respect to $\theta$ when (a) $|a|>|b|$ and (b) $|b|>|a|$.
        (a) Continuous (red) curve shows $\dot \psi$ when $a>b>0$, and dashed (blue) curve shows $\dot \psi$ when $a<0$ and $b>0$.
        (b) Continuous (red) curve shows $\dot \psi$ when $b>a>0$, and dashed (blue) curve shows $\dot \psi$ when $a>0$ and $b<0$.
        }
        \label{fig:cdps}
        \end{center}
\end{figure}

Now, we will analyze the situation when $|a|>|b|$.
The numerator, $\cos^2 \theta-\frac{2b}{a} \cos \theta+1$, is the same as the terms inside the parentheses in equation \eqref{derphidot}, and we have seen that there is not any root for these terms when $|a|>|b|$. 
For the interval $\theta\in (0,\pi/2)$, $1+\cos^2 \theta-2 b/a \cos \theta > (1-\cos \theta)^2$, then the numerator is positive in this interval.
This situation together with the absence of root shows that the numerator is always positive.
The positiveness of numerator together with positiveness of $\sin^2 \theta$ and $\epsilon$ shows that the result of terms inside the parentheses in equation \eqref{psidot2} is always positive.
This means that $\dot \psi$ has always the same sign as $a$ when $|a|>|b|$.
Therefore, $\dot \psi$ goes to $sgn(a)\infty$ as $\theta$ goes to $0$ or $\pi$ since $\sin^2 \theta$ goes to $0$, and $\dot \psi$ does not have any root and spin reversal is impossible when $|a|>|b|$.
One can see these changes for two possible sign combination of $a$ and $b$ in figure \ref{fig:cdps}(a).

Now, we will analyze the situation when $|b|>|a|$.
The numerator in equation \eqref{psidot2} has a root when $|b|>|a|$, and then the result of the terms in the numerator can be positive or negative.
The numerator in equation \eqref{psidot2} goes to $2(1-b/a)$ as $\theta$ goes to $0$, and it goes to $2(1+b/a)$ as $\theta$ goes to $\pi$.
The denominator, $\sin^2 \theta$, goes to $0$ in these limits, then the first term in equation \eqref{psidot2} goes to infinity in one limit and minus infinity in the other one depending on the sign of $b/a$ when $|b|>|a|$.
By multiplying these limits with $a$, which is present in front of the parentheses in equation \eqref{psidot2}, we see that $\dot \psi$ goes to $-sgn(b)\infty$ as $\theta$ goes to $0$ and $sgn(b)\infty$ as $\theta$ goes to $\pi$.
Then, there is always a root of $\dot \psi$ when $|b|>|a|$.
One can see from equation \eqref{psidot2} that when $b/a>1$ the root is between $0$ and $\pi/2$; and when $b/a<-1$ the root is between $\pi/2$ and $\pi$.
One can see two examples of this situation in figure \ref{fig:cdps}(b).

After some rearrangements in equation \eqref{psidot}, one can write it as $\dot \psi=a[(1-I_x/I_z) \cos^2 \theta-(b/a) \cos \theta+I_x/I_z]/\sin^2 \theta$, and then the root or the angle where spin reversal takes place when $|b|>|a|$ can be found as
\begin{equation}
	\theta_{r_{1,2}}=\arccos\left[ \frac{ \frac{b}{a}\pm\sqrt{ \frac{b^2}{a^2}-4 \frac{I_x}{I_z}\left(1-\frac{I_x}{I_z}\right)} }{2\left(1-\frac{I_x}{I_z} \right)}\right].  \tag{\ref{rootdotpsi}}
\end{equation}
Either plus or minus sign gives the angle, not both.

The derivative of $\dot \psi(\theta)$ can be obtained as
\begin{equation}
	\frac{d \dot \psi (\theta)}{d \theta}=\frac{b}{\sin^3 \theta}\left[ 1+\cos^2 \theta-\frac{2a}{b} \cos \theta \right],
\end{equation}
and then, turning points can be obtained as
\begin{equation}
	\theta_{tp}=\arccos \left( \frac{a}{b}\pm \sqrt{\frac{a^2}{b^2}-1}\right).
\end{equation}
This shows that the turning point is available only if $|a|>|b|$, which is consistent with the above considerations.
The turning point is between $0$ and $\pi/2$ when $a/b>1$, and it is between $\pi/2$ and $\pi$ when $a/b<-1$.
One can see these turning points in figure \ref{fig:cdps}(a) for two possible sign combinations for $a$ and $b$.

$\dot \psi$ can either increase or decrease, or can both increase and decrease in a half nutation period when $|a|>|b|$; and it can either increase or decrease in a half nutation period and can change sign when $|b|>|a|$.
To see what will happen, one should consider the possible interval of $\theta$ and the relation between $a$ and $b$.

\section{Appendix 3}
\label{ap3}

In this part, we will give the relations between quantities used in this work and parameters used by Routh \cite{Routh}.
We will also compare the figures given by Routh with the results of this work.
To avoid confusion between the parameters, we will designate parameters and quantities used by Routh in square parentheses.

The correspondences between symbols representing moments of inertia are $[A]=I_x$, $[B]=I_y$ and $[C]=I_z$, for symmetric top $[A]=[B]$.
The relations between angular velocities are as follows:
\begin{eqnarray}
	[w_3] &=& [n] =\dot \psi+\dot \phi \cos \theta, \\
	{[}-w_1] &=& [\dot \Psi \sin \theta]= \dot \phi \sin \theta, \\
	{[}w_2] &=& \dot \theta.
\end{eqnarray}

In equation (2) of Art. 200, Routh has given the relation $[-A w_1 \sin \theta+ C n \cos \theta]=[E]$ which can be written as $I_x \dot \phi \sin^2 \theta+I_z (\dot \psi+\dot \phi \cos \theta)\cos \theta$, and this is equal to $L_{z'}=I_x b$, i.e. equation \eqref{angmom}, then in short $[E]=I_x b$.
In equation (3), Routh has given the relation $[A(w_1^2+w_2^2)+C n^2=F-2gh \cos \theta]$ and the mass of the top is taken as unity.
From this relation, it can be obtained that $[F]=2E$, which is twice the energy given by equation \eqref{enrgy}.

There are four parameters in Routh's work to determine different types of motion (Art. 204): $[l]$, $[c]$, $[a]$ and $[OQ]$.
The length $[l]=[A/h]$ used by Routh can be written in terms of quantities used in this work as $[l]=I_x/l$.
The second parameter is given by $[c]=[l^2\left( C(w_1^2+w_2^2)+2 g h w_1 \sin \theta/ n\right)/(2Cg)]$ and it can be obtained as
\begin{equation}
	[c]=\frac {I_x}{l} \frac {1}{Mgl} \left(E'-Mgl \frac{b}{a}\right).
	\label{parac}
\end{equation}
This parameter corresponds to the relation between $E'$ and $Mgl b/a$ which is used above.

The third parameter is given by $[a]=[E l/(Cn)]$ which can be obtained as
\begin{equation}
	[a]= \frac{I_x}{l} \frac{b}{ a}.
	\label{paraa}
\end{equation}
It can be seen that this parameter corresponds to the ratio of conserved angular momenta.

The last parameter gives a bit of a strange result in terms of the quantities used in this work.
From the last lines of Art. 204 b, it can be inferred that $[OQ^2]$ is equivalent to $[(a+2c)^2-8plc]$.
The parameters in this relation are already given in terms of quantities of this work except $[p]$, and from the explanation after equation (2) in Art. 204, it can be seen that $[p]=[C^2 n^2/(4 g h^2 l)]$ which can be written in terms of quantities of this work as $[p]=I_x a^2/(4 Mgl)$.
Then, one can obtain
\begin{equation}
	[OQ^2]=\left( \frac{I_x}{l}\right)^2 \Bigg\{  \frac{b^2}{a^2}+\frac{4}{(Mgl)^2} \left( E'-Mgl \frac{b}{a}\right)\left(E'-\frac{I_x a^2}{2}\right) \Bigg\}.
	\label{paraoq}
\end{equation}
This relation is not simplifiable similar to other ones.

Now, we will compare Routh's classification and the results of this work.
The classification in his work is done firstly with respect to $[c]$.
From equation \eqref{parac}, it can be seen that $[c]>0$ (Art. 204 a), $[c]<0$ (Art. 204 b) and $[c]=0$ (Art. 204 c) correspond to $E'>Mgl b/a$, $E'<Mgl b/a$ and $E'=Mgl b/a$, respectively.
After classification with respect to $[c]$, the relation between $[l]$ and $[a]$ or $[l]$ and $[OQ]$ is considered in Routh's work.
Comparison done by using $[l]$ and $[a]$ corresponds to comparison of this work by using $a$ and $b$ which can be understood from equation \eqref{paraa} and $[l]=I_x/l$.
However, the usage of $[l]$ and $[OQ]$ does not correspond to anything used in this work.
Then, we will leave the comparison of cases related to $[l]$ and $[OQ]$ (Art. 204 b) to the end.

For the first figure of Art. 204 a, $[l]>[a]$ and $[c>0]$ which is equivalent to $a>b$ and $E'>Mgl b/a$.
This case corresponds to looping motion, and from section \ref{loopingmotion}, it can be seen that the classification and figure are consistent with Routh's work.
For the first figure of Art. 204 c, $[l]>[a]$ and $[c]=0$ which is equivalent to $a>b$ and $E'=Mgl b/a$.
From section \ref{mwc}, one can see that this case gives motion with cusps and it can be seen that figures and classifications are consistent.

We have mentioned in section \ref{bgraterthana} that when $|b|>|a|$, the relation between $E'$ and $Mglb/a$ does not affect shapes for the locus similar to cases satisfying $|a|>|b|$.
This can also be seen from the second figures of Art. 204 a and Art. 204 c, for whom $[a]>[l]$ or $|b|>|a|$.
We should note that when $|b|>|a|$, the precession is more than $2 \pi$ in one nutation period in numerical solutions, 
and in figures given by Routh, the precession is less than $2 \pi$ in one nutation period when $[a]>[l]$.
Other than this small difference, it can also be seen that the second figures of Art. 204 a and Art. 204 c are similar to each other and they are also similar to the result of section \ref{mwepavam}, where $b>a$.
Then, one can say that Routh's classification and figures are similar to the ones in this work when $[a]>[l]$ or $|b|>|a|$.

The explanations for the third figures of Art. 204 a and Art. 204 c are given as $[l]=[a]$ which corresponds to $b=a$ and in this case $Mglb/a$ becomes $Mgl$.
The situation $|b|=|a|$ is studied previously, and the third figure of Art. 204 a can correspond to either "motion of a fast top" or "motion over the bump" \cite{Tanriverdi2020}.
Both cases give similar results to the third figure of Art. 204 a where $[c]>0$ or $E'>Mgl$ in both cases.
The third figure of Art. 204 c where $[l]=[a]$ and $[c]=0$ corresponds to $b=a$ when $E'=Mgl$, and this case can give "spiraling motion" for a slow top and the result of previous study \cite{Tanriverdi2020} is consistent with Routh's work.
But, we should note that a top satisfying $b=a \ge \sqrt{4 Mgl/I_x}$ and $E'=Mgl$ corresponds to sleeping top.

Now, we will compare cases where $[l]$ and $[OQ]$ are used in Routh's work, Art. 204 b, where $[c]<0$ or $E'<Mglb/a$.
We have already mentioned that $[OQ]$ is not simplifiable similar to other parameters.
Comparison $[l]$ and $[OQ]$ is obtained from squares of these quantities in Routh's work, and we will use their square for comparison.
$[l^2]$ is equal to $I_x^2/l^2$ in terms of parameters of this work and $[OQ^2]$ is given in equation \eqref{paraoq}.
From that equation, one can see that it is different from our classification since $[OQ^2]$ does not depend on only $b$ and $a$. 
Then, we will consider different cases satisfying $[c]<0$ or $E'<Mglb/a$ relation and look at whether given conditions for figures in Art. 204 b are held or not.
Previously, we did not specify the value of $l$ for the considered heavy symmetric top or gyroscope, and it will be taken as $l=3.13 \times 10^{-2}\,m$.
Then, one obtains $[l^2]$ or $I_x^2/l^2$ as $5.31 \times 10^{-5} \,kg^2 \,m^2$.

The first figure in Art. 204 b gives a case that precession angular velocity $\dot \phi$ has the same sign at both extrema of $\theta$ and there is more than one nutation period in one precession period.
We have studied such a case: wavy precession, section \ref{wp}, in which $E'<Mglb/a$ and $a>b$.
We have mentioned that when $a>b$ and $E'<Mglb/a$, $\dot \phi$ never becomes zero or changes sign and there is more than one nutation period in one precession period.
$E'<Mglb/a$ is consistent with the condition of Art. 204 b, i.e. $[c]<0$.
If one calculates $[OQ^2]$ for the case in section \ref{wp}, one obtains $5.18 \times 10^{-5} \,kg^2 \,m^2$ which is smaller than $[l^2]$ and consistent with the explanation given in the first figure in Art. 204 b.
Then, we need to find a case that does not satisfy Routh's condition and gives a similar motion to the shown one in the first figure of Art. 204 b.

\begin{figure}[!h]
       \begin{center}
       \subfigure[]{
	       \includegraphics[width=4.2cm]{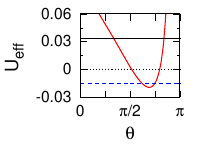}
       }
       \subfigure[]{
               \includegraphics[width=4.2cm]{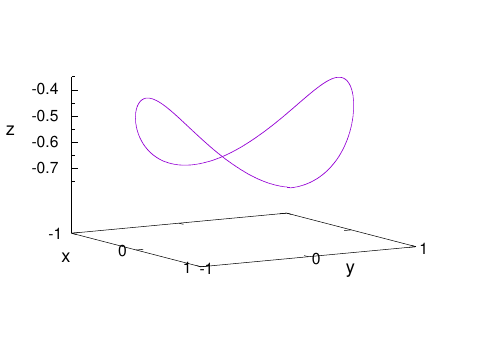}
       }
       \caption{(a) $U_{eff}$ (red curve), $E'$ (dashed blue line) and $Mgl b/a$ (black line).
                (b) Shapes for the locus on the unit sphere when $a>b$ for evaluation of parameter $[OQ]$. 
                Initial values are $\theta_0=2.35 \,rad$, $\dot \theta_0=0$, $\dot \phi_0=24.0 \,rad \,s^{-1}$ and $\dot \psi_0=56.8 \,rad \,s^{-1}$, and constants are $a=10.0 \,rad \,s^{-1}$, $b=5.0 \,rad \,s^{-1}$, $E'=-0.0150 \,J$ and $Mgl b/a=0.0340 \,J$.
                Turning angles are $\theta_{max}=2.35 \,rad$ and $\theta_{min}=1.95 \,rad$.
                The animated plot is available at \href{https://youtu.be/Qjsa_E354S8}{https://youtu.be/Qjsa\_E354S8}. 
	       }
       \label{fig:uefftt_routh_agtb}
       \end{center}
\end{figure}

\begin{figure}[!h]
	        \begin{center}
		\subfigure[$\theta$]{
			\includegraphics[width=3.7cm]{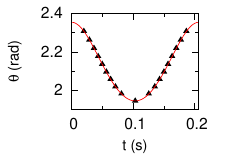}
		}
		\subfigure[$\phi$]{
			\includegraphics[width=3.7cm]{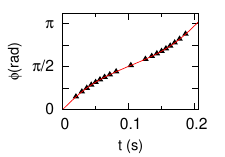}
		}
		\subfigure[$\psi$]{
			\includegraphics[width=3.7cm]{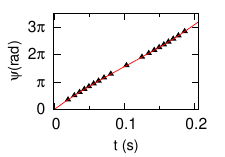}
		}
		\subfigure[$\dot \theta$]{
			\includegraphics[width=3.7cm]{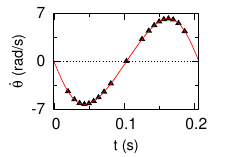}
		}
		\subfigure[$\dot \phi$]{
			\includegraphics[width=3.7cm]{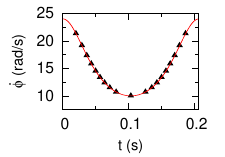}
		}
		\subfigure[$\dot \psi$]{
			\includegraphics[width=3.7cm]{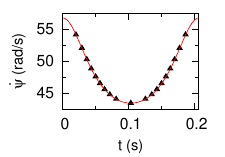}
		}
			\caption{(a) $\theta$, (b) $\phi$, (c) $\psi$, (d) $\dot \theta$, (e) $\dot \phi$ and (f) $\dot \psi$ for the evaluation of parameter $[OQ]$ when $a>b$.
			Initial values are same with figure \ref{fig:uefftt_routh_agtb}.
			}
		\label{fig:ttdfd_routh_agtb}
		\end{center}
\end{figure}

A case satisfying $a=10.0 \,rad \,s^{-1}$, $b=5.0 \,rad \,s^{-1}$, $E'=-0.0150 \,J$ and $Mgl b/a=0.0340 \,J$ is considered as the first example.
$[OQ^2]$ is equal to $7.26 \times 10^{-5} \,kg^2 \,m^2$ for this case, and it is greater than $[l^2]$, which is inconsistent with the explanation of the first figure in Art. 204 b.
On the other hand, this example safisfies $E'<Mglb/a$ and $a>b$ conditions.
Results of the numerical calculations for this case can be seen in figures \ref{fig:uefftt_routh_agtb} and \ref{fig:ttdfd_routh_agtb}.
One can see from figure \ref{fig:ttdfd_routh_agtb}(e) that $\dot \phi$ does not change sign, and there are more than one nutation period in one precession period, which can be seen in figure \ref{fig:uefftt_routh_agtb}(b). 

One may consider that the possible interval of $\theta$ is between $\pi/2$ and $\pi$ in the mentioned example, and this example is different from the case given in the first figure of Art. 204 b.
It is also possible to obtain a case that $\theta$ is always smaller than $\pi/2$ and $\dot \phi$ has the same sign at both extrema while nutation period is smaller than precession period by taking initial values as $\theta_0=1.49 \,rad$, $\dot \theta_0=0$, $\dot \phi_0=18.5 \,rad \,s^{-1}$ and $\dot \psi_0=82.3 \,rad \,s^{-1}$.
This time the constants can be obtained as $a=21.0 \,rad \,s^{-1}$, $b=20.0 \,rad \,s^{-1}$, $E'=0.0440 \,J$ and $Mgl b/a=0.06476 \, J$, and extrema of $\theta$ become $\theta_{max}=1.49 \, rad$ and $\theta_{min}=1.26 \, rad$.
In this case, $[OQ^2]$ is equal to $5.41 \times 10^{-5} \,kg^2 \,m^2$ which is again greater than $[l^2]$ and contradicts with the explanation given for the first figure of Art. 204 b.
The previous case is chosen to show the resemblance to wavy precession because in this case the precession period is a bit longer than the nutation period and the resemblance is less.
Nevertheless, these two cases show that Routh's explanation given for the first figure of Art. 204 b is not always valid.

\begin{figure}[!h]
      \begin{center}
      \subfigure[]{
	      \includegraphics[width=4.2cm]{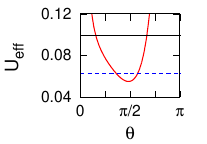}
      }
      \subfigure[]{
              \includegraphics[width=4.2cm]{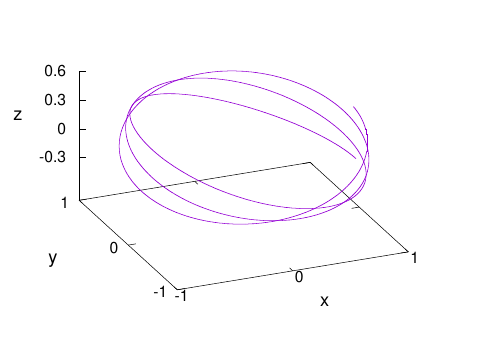}
      }
      \caption{(a) $U_{eff}$ (red curve), $E'$ (dashed blue line) and $Mgl b/a$ (black line).
               (b) Shapes for the locus on the unit sphere when $b>a$ for evaluation of parameter $[OQ]$.
               Initial values are $\theta_0=1.81 \,rad$, $\dot \theta_0=0$, $\dot \phi_0=27.0 \,rad \,s^{-1}$ and $\dot \psi_0=66.2 \,rad \,s^{-1}$, and constants are $a=15.0 \,rad \,s^{-1}$, $b=22.0 \,rad \,s^{-1}$, $E'=0.0627 \,J$ and $Mgl b/a=0.0997 \,J$.
               Turning angles are $\theta_{max}=1.81 \,rad$ and $\theta_{min}=1.14 \,rad$.
               The animated plot is available at \href{https://youtu.be/hCwuXnifHzs}{https://youtu.be/hCwuXnifHzs}.
      }
      \label{fig:uefftt_routh_astb}
      \end{center}
\end{figure}

The second figure of Art. 204 b is similar to the second figures of Art. 204 a and Art. 204 c, and condition for these two cases is $[l]<[a]$ which is equivalent to $|b|>|a|$.
One of the main properties of cases when $|b|>|a|$ is that the nutation period is greater than the precession period.
We have already mentioned that shapes for the locus when $|b|>|a|$ are similar to each other, and one can not easily distinguish one from another similar to the cases when $|a|>|b|$.
We have also mentioned that when $|b|>|a|$, $Mglb/a$ can be smaller than the minimum of $U_{eff}$ and one can not always find a case satisfying $E'<Mglb/a$ or $[c]<0$ relation.
To get cases satisfying $E'<Mglb/a$ relation when $|b|>|a|$, one should choose smaller values of $a$ and $b$.

\begin{figure}[!h]
                \begin{center}
			\subfigure[$\theta$]{                                                                                                                                                                                   \includegraphics[width=3.7cm]{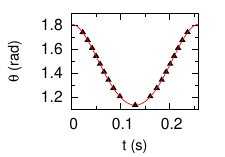}
                }       
                \subfigure[$\phi$]{
	                \includegraphics[width=3.7cm]{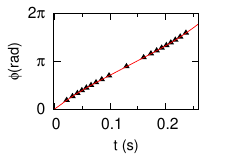}
                }
                \subfigure[$\psi$]{
                        \includegraphics[width=3.7cm]{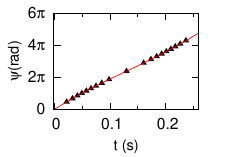}
                }       
                \subfigure[$\dot \theta$]{
                        \includegraphics[width=3.7cm]{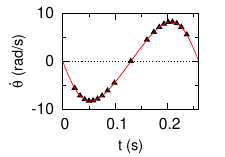}
                }       
                \subfigure[$\dot \phi$]{
                       \includegraphics[width=3.7cm]{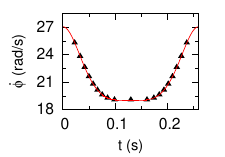}
                }       
                \subfigure[$\dot \psi$]{
                       \includegraphics[width=3.7cm]{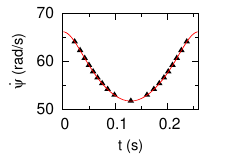}
                }       
			\caption{(a) $\theta$, (b) $\phi$, (c) $\psi$, (d) $\dot \theta$, (e) $\dot \phi$ and (f) $\dot \psi$ for the evaluation of parameter $[OQ]$ when $a<b$.
                        Initial values are same with figure \ref{fig:uefftt_routh_astb}. 
                        }
                \label{fig:ttdfd_routh_astb}
                \end{center}
\end{figure}

One can find many examples satisfying both $|b|>|a|$ and $[l^2]<[OQ^2]$ relations when $E'<Mglb/a$ or $[c]<0$.
But to show inconsistency, one needs to find a contradictory case.
As a contradictory example to the second figure of Art. 204 b, a case satisfying $a=15.0 \,rad \,s^{-1}$, $b=22.0 \,rad \,s^{-1}$, $E'=0.0627 \,J$ and $Mgl b/a=0.0997 \,J$ will be considered. 
One can see that $E'<Mglb/a$ and $b>a$.
In this case, $[l^2]$ is again the same, and $[OQ^2]$ is equal to $ 5.12 \times 10^{-5} \,kg^2 \,m^2$, which is smaller than $[l^2]$, and this situation also contradicts with the explanation given in the second figure of Art. 204 b.
Results of the numerical solutions of this case can be seen in figures \ref{fig:uefftt_routh_astb} and \ref{fig:ttdfd_routh_astb}, and shapes for the locus are naturally similar to cases satisfying $|b|>|a|$ relation.
Therefore, this case shows that Routh's explanation given for the second figure of Art. 204 b is not always valid.

It has been seen that one can obtain a case satisfying the explanation of the first figure of Art. 204 b but giving a motion similar to the second figure of Art. 204 b; and a case satisfying the explanation of the second figure of Art. 204 b but giving motion similar to the first figure of Art. 204 b.
This shows that comparing $[OQ]$ with $[l]$ is inadequate.

We should note that if $[OQ]$ is replaced with $[a]$ in Art. 204 b, the comparison becomes equivalent to the comparison between $a$ and $b$, 
and then the contradiction and problematic situations disappear.

We should also note that there is another possible motion when $[c]<0$ which is not mentioned in Art. 204 b.
When $|b|=|a|$, $Mglb/a$ becomes $Mgl$ and the related relation becomes $E'<Mgl$.
One can find such a case in previous work \cite{Tanriverdi2020}.

\section{Appendix 4}
\label{ap4}

Results of numerical solutions for $\theta$, $\phi$, $\psi$, $\dot \theta$, $\dot \phi$ and $\dot \psi$ for the studied cases can be seen below.
Initial values of $\phi$ and $\psi$ are taken as zero.
For the numerical solution including constants, $\phi$ and $\psi$ are calculated by using
\begin{eqnarray}
	\phi&=& \int \frac{\dot \phi} {\dot \theta} d \theta, \nonumber \\
	\psi&=& \int \frac{\dot \psi} {\dot \theta} d \theta.
\end{eqnarray}
Uneven time interval in the second method is result of integrating with respect to $\theta$.

\begin{figure}[!h]
	\begin{center}
		\subfigure[$\theta$]{
			\includegraphics[width=3.7cm]{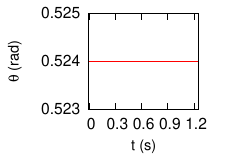}
		}
		\subfigure[$\phi$]{
			\includegraphics[width=3.7cm]{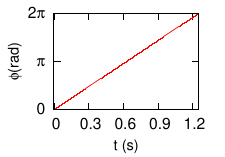}
		}
		\subfigure[$\psi$]{
			\includegraphics[width=3.7cm]{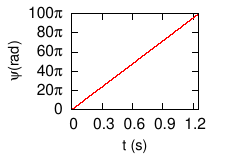}
		}
		\subfigure[$\dot \theta$]{
			\includegraphics[width=3.7cm]{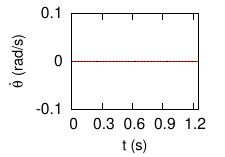}
		}
		\subfigure[$\dot \phi$]{
			\includegraphics[width=3.7cm]{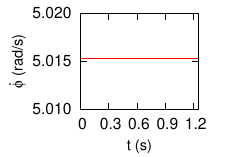}
		}
		\subfigure[$\dot \psi$]{
			\includegraphics[width=3.7cm]{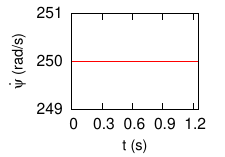}
		}
		\caption{(a) $\theta$, (b) $\phi$, (c) $\psi$, (d) $\dot \theta$, (e) $\dot \phi$ and (f) $\dot \psi$ for the regular precession when $|a|>|b|$.
		Initial values are same with figure \ref{fig:uefftt_7a1}. 
		}
		\label{fig:ttdfd_7a1}
	\end{center}
\end{figure}

\begin{figure}[!h]
	\begin{center}
		\subfigure[$\theta$]{
			\includegraphics[width=3.7cm]{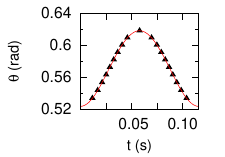}
		}
		\subfigure[$\phi$]{
			\includegraphics[width=3.7cm]{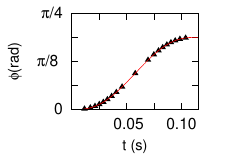}
		}
		\subfigure[$\psi$]{
			\includegraphics[width=3.7cm]{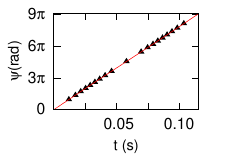}
		}
		\subfigure[$\dot \theta$]{
			\includegraphics[width=3.7cm]{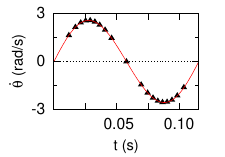}
		}
		\subfigure[$\dot \phi$]{
			\includegraphics[width=3.7cm]{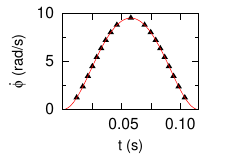}
		}
		\subfigure[$\dot \psi$]{
			\includegraphics[width=3.7cm]{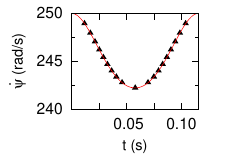}
		}
		\caption{(a) $\theta$, (b) $\phi$, (c) $\psi$, (d) $\dot \theta$, (e) $\dot \phi$ and (f) $\dot \psi$ for the motion with cusps.
		Curves (red) show results obtained from the integration of angular accelerations, and (black) triangles show the results obtained from the second numerical solution.
		Initial values are same with figure \ref{fig:uefftt_1}.
		}
		\label{fig:ttdfd_1} 
	\end{center}
\end{figure}

\begin{figure}[!h]
	\begin{center}
		\subfigure[$\theta$]{
			\includegraphics[width=3.7cm]{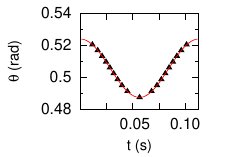}
		}
		\subfigure[$\phi$]{
			\includegraphics[width=3.7cm]{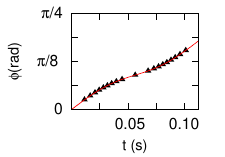}
		}
		\subfigure[$\psi$]{
			\includegraphics[width=3.7cm]{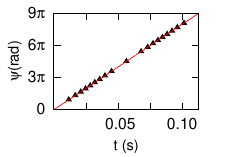}
		}
		\subfigure[$\dot \theta$]{
			\includegraphics[width=3.7cm]{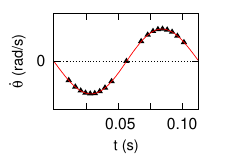}
		}
		\subfigure[$\dot \phi$]{
			\includegraphics[width=3.7cm]{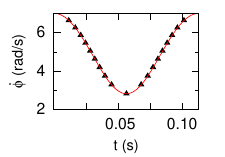}
		}
		\subfigure[$\dot \psi$]{
			\includegraphics[width=3.7cm]{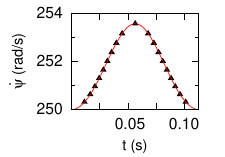}
		}
		\caption{(a) $\theta$, (b) $\phi$, (c) $\psi$, (d) $\dot \theta$, (e) $\dot \phi$ and (f) $\dot \psi$ for wavy precession.
		Legends are the same with figure \ref{fig:ttdfd_2}.
		Initial values are same with figure \ref{fig:uefftt_2}.
		}
		\label{fig:ttdfd_2}
	\end{center}
\end{figure}

\begin{figure}[!h]
	\begin{center}
		\subfigure[$\theta$]{
			\includegraphics[width=3.7cm]{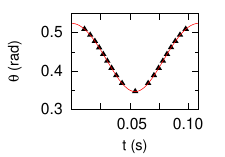}
		}
		\subfigure[$\phi$]{
			\includegraphics[width=3.7cm]{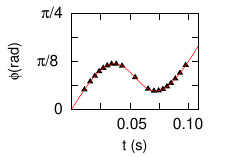}
		}
		\subfigure[$\psi$]{
			\includegraphics[width=3.7cm]{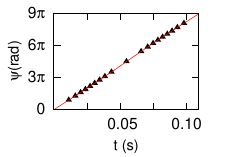}
		}
		\subfigure[$\dot \theta$]{
			\includegraphics[width=3.7cm]{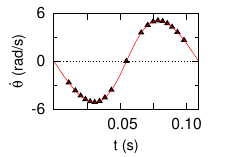}
		}
		\subfigure[$\dot \phi$]{
			\includegraphics[width=3.7cm]{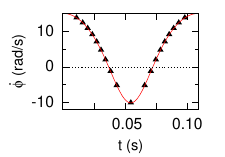}
		}
		\subfigure[$\dot \psi$]{
			\includegraphics[width=3.7cm]{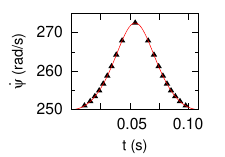}
		}
		\caption{(a) $\theta$, (b) $\phi$, (c) $\psi$, (d) $\dot \theta$, (e) $\dot \phi$ and (f) $\dot \psi$ for looping motion.
		Legends are the same with figure \ref{fig:ttdfd_2}.
		Initial values are same with figure \ref{fig:uefftt_3}.
		}
		\label{fig:ttdfd_3}
	\end{center}
\end{figure}

\begin{figure}[!h]
	\begin{center}
		\subfigure[$\theta$]{
			\includegraphics[width=3.7cm]{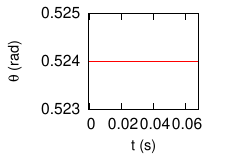}
		}
		\subfigure[$\phi$]{
			\includegraphics[width=3.7cm]{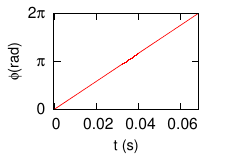}
		}
		\subfigure[$\psi$]{
			\includegraphics[width=3.7cm]{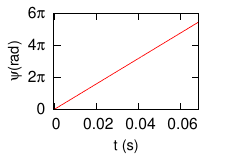}
		}
		\subfigure[$\dot \theta$]{
			\includegraphics[width=3.7cm]{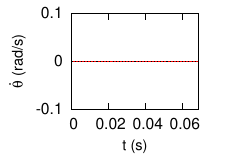}
		}
		\subfigure[$\dot \phi$]{
			\includegraphics[width=3.7cm]{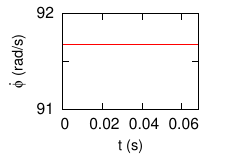}
		}
		\subfigure[$\dot \psi$]{
			\includegraphics[width=3.7cm]{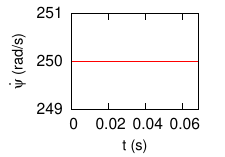}
		}
		\caption{(a) $\theta$, (b) $\phi$, (c) $\psi$, (d) $\dot \theta$, (e) $\dot \phi$ and (f) $\dot \psi$ for the regular precession when $|b|>|a|$.
		Initial values are same with figure \ref{fig:uefftt_7a2}.
		}
		\label{fig:ttdfd_7a2}
	\end{center}
\end{figure}

\begin{figure}[!h]
	\begin{center}
		\subfigure[$\theta$]{
			\includegraphics[width=3.7cm]{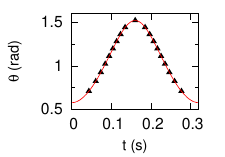}
		}
		\subfigure[$\phi$]{
			\includegraphics[width=3.7cm]{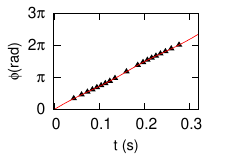}
		}
		\subfigure[$\psi$]{
			\includegraphics[width=3.7cm]{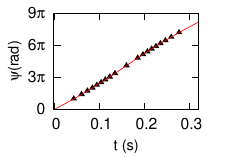}
		}
		\subfigure[$\dot \theta$]{
			\includegraphics[width=3.7cm]{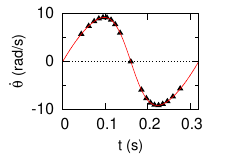}
		}
		\subfigure[$\dot \phi$]{
			\includegraphics[width=3.7cm]{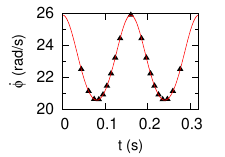}
		}
		\subfigure[$\dot \psi$]{
			\includegraphics[width=3.7cm]{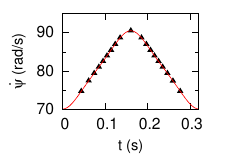}
		}
		\caption{(a) $\theta$, (b) $\phi$, (c) $\psi$, (d) $\dot \theta$, (e) $\dot \phi$ and (f) $\dot \psi$ for motion with the same precession angular velocity at both extrema.
		Legends are the same with figure \ref{fig:ttdfd_2}.
		Initial values are same with figure \ref{fig:uefftt_4}.
		}
		\label{fig:ttdfd_4}
	\end{center}
\end{figure}

\begin{figure}[!h]
	\begin{center}
		\subfigure[$\theta$]{
			\includegraphics[width=3.7cm]{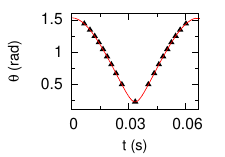}
		}
		\subfigure[$\phi$]{
			\includegraphics[width=3.7cm]{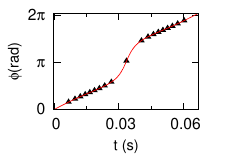}
		}
		\subfigure[$\psi$]{
			\includegraphics[width=3.7cm]{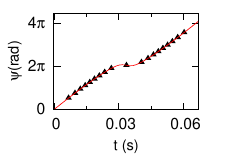}
		}
		\subfigure[$\dot \theta$]{
			\includegraphics[width=3.7cm]{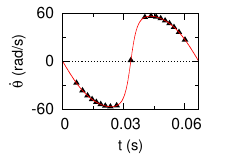}
		}
		\subfigure[$\dot \phi$]{
			\includegraphics[width=3.7cm]{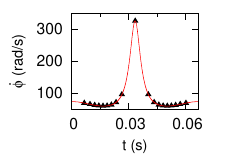}
		}
		\subfigure[$\dot \psi$]{
			\includegraphics[width=3.7cm]{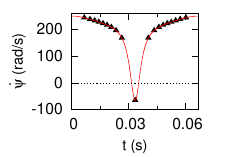}
		}
		\caption{(a) $\theta$, (b) $\phi$, (c) $\psi$, (d) $\dot \theta$, (e) $\dot \phi$ and (f) $\dot \psi$ for spin reversing motion when $|b|>|a|$.
		Legends are the same with figure \ref{fig:ttdfd_2}.
		Initial values are same with figure \ref{fig:uefftt_sr_b1}.
		}
		\label{fig:ttdfd_sr_b1}
	\end{center}
\end{figure}

\begin{figure}[!h]
	\begin{center}
		\subfigure[$\theta$]{
			\includegraphics[width=3.7cm]{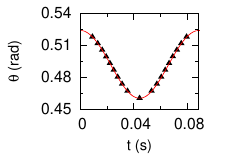} 
		}
		\subfigure[$\phi$]{
			\includegraphics[width=3.7cm]{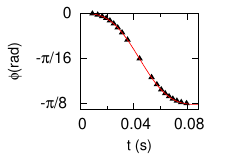}
		}
		\subfigure[$\psi$]{
			\includegraphics[width=3.7cm]{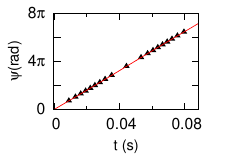}
		}
		\subfigure[$\dot \theta$]{
			\includegraphics[width=3.7cm]{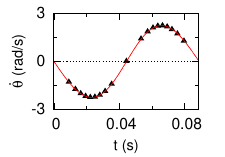}
		}
		\subfigure[$\dot \phi$]{
			\includegraphics[width=3.7cm]{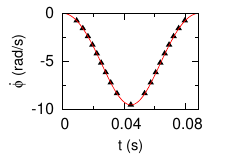}
		}
		\subfigure[$\dot \psi$]{
			\includegraphics[width=3.7cm]{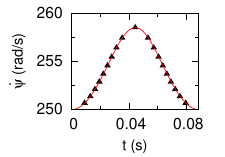}
		}
		\caption{(a) $\theta$, (b) $\phi$, (c) $\psi$, (d) $\dot \theta$, (e) $\dot \phi$ and (f) $\dot \psi$ for negative $Mgl$.
		Legends are the same with figure \ref{fig:ttdfd_2}.
		Initial values are same with figure \ref{fig:uefftt_mmgl}.
		}
		\label{fig:ttdfd_mmgl}
	\end{center}
\end{figure}

\end{document}